\documentclass[twocolumn]{aastex631}

\usepackage{graphicx}	
\usepackage{amsmath}	
\usepackage{xcolor}
\usepackage{threeparttable}
\usepackage{url}
\usepackage{hyperref}
\usepackage{booktabs}
\usepackage{bm}
\usepackage{newtxtext,newtxmath}

\newcommand{\ang}{\text{\AA}}
\newcommand{\msun}{{\rm M}_\odot}

\newcommand{\vect}[1]{\boldsymbol{#1}}
\newcommand{\oii}{[\ion{O}{2}]\,$\lambda\lambda\,3727,3729$ }
\newcommand{\oiii}{[\ion{O}{3}]\,$\lambda\,5008$ }

\newcommand{\hii}{\ion{H}{2} }

\defcitealias{Chen2023}{Paper I}

\shorttitle{Turbulence in QSO nebulae}
\shortauthors{Chen et al.}
\graphicspath{{./}{figures/}}

\begin{document}

\title{An ensemble study of turbulence in extended QSO nebulae at \bm{$z\approx 0.5$}--1}

\author[0000-0002-8739-3163]{Mandy C. Chen}
\affiliation{Department of Astronomy and Astrophysics, The University of Chicago, Chicago, IL 60637, USA}

\author[0000-0001-8813-4182]{Hsiao-Wen Chen}
\affiliation{Department of Astronomy and Astrophysics, The University of Chicago, Chicago, IL 60637, USA}

\author[0000-0002-1690-3488]{Michael Rauch}
\affiliation{The Observatories of the Carnegie Institution for Science, 813 Santa Barbara Street, Pasadena, CA 91101, USA}

\author[0000-0002-2941-646X]{Zhijie Qu}
\affiliation{Department of Astronomy and Astrophysics, The University of Chicago, Chicago, IL 60637, USA}

\author[0000-0001-9487-8583]{Sean D. Johnson}
\affiliation{Department of Astronomy, University of Michigan, Ann Arbor, MI 48109, USA}

\author[0000-0002-0668-5560]{Joop Schaye}
\affiliation{Leiden Observatory, Leiden University, PO Box 9513, NL-2300 RA Leiden, the Netherlands}

\author[0000-0002-8459-5413]{Gwen C. Rudie}
\affiliation{The Observatories of the Carnegie Institution for Science, 813 Santa Barbara Street, Pasadena, CA 91101, USA}

\author[0000-0002-0311-2812]{Jennifer I-Hsiu Li}
\affiliation{Department of Astronomy, University of Michigan, Ann Arbor, MI 48109, USA}
\affiliation{Michigan Institute for Data Science, University of Michigan, Ann Arbor, MI, 48109, USA}

\author[0000-0002-2662-9363]{Zhuoqi (Will) Liu}
\affiliation{Department of Astronomy, University of Michigan, Ann Arbor, MI 48109, USA}

\author[0000-0001-7869-2551]{Fakhri S. Zahedy}
\affiliation{The Observatories of the Carnegie Institution for Science, 813 Santa Barbara Street, Pasadena, CA 91101, USA}

\author[0000-0001-5804-1428]{Sebastiano Cantalupo}
\affiliation{Department of Physics, University of Milan Bicocca, Piazza della Scienza 3, I-20126 Milano, Italy}

\author[0000-0003-3244-0409]{Erin Boettcher}
\affiliation{Department of Astronomy, University of Maryland, College Park, MD 20742, USA}
\affiliation{X-ray Astrophysics Laboratory, NASA/GSFC, Greenbelt, MD 20771, USA}
\affiliation{Center for Research and Exploration in Space Science and Technology, NASA/GSFC, Greenbelt, MD 20771, USA}



\begin{abstract}

Turbulent motions in the circumgalactic medium (CGM) play a critical role in regulating the evolution of galaxies, yet their detailed characterization remains elusive. Using two-dimensional velocity maps constructed from spatially-extended [\ion{O}{2}] and [\ion{O}{3}] emission,
\cite{Chen2023} measured the velocity structure functions (VSFs) of 
four quasar nebulae at $z\approx\!0.5$--1.1. One of these 
exhibits a spectacular Kolmogorov relation.  
Here we carry out an ensemble study 
using an expanded sample incorporating four new nebulae from three additional QSO fields.  
The VSFs measured for all eight nebulae are best explained by subsonic turbulence revealed by the line-emitting gas, which in turn strongly suggests that 
the cool gas ($T\!\sim\!10^4$ K) is dynamically coupled to the hot ambient medium. 
Previous work demonstrates that the largest nebulae in our sample reside in group environments with clear signs of tidal interactions, suggesting that environmental effects are vital in seeding and enhancing turbulence within the gaseous halos, ultimately promoting the formation of the extended nebulae. No discernible differences are observed in the VSF properties between radio-loud and radio-quiet QSO fields. We estimate the turbulent heating rate per unit volume, $Q_{\rm turb}$, in the QSO nebulae to be $\sim 10^{-26}$--$10^{-22}$ erg cm$^{-3}$ s$^{-1}$ for the cool phase and $\sim 10^{-28}$--$10^{-25}$ erg cm$^{-3}$ s$^{-1}$ for the hot phase. This range aligns with measurements in the intracluster medium and star-forming molecular clouds but is $\sim10^3$ times higher than the $Q_{\rm turb}$ observed 
inside cool gas clumps 
on scales $\lesssim1$ kpc using absorption-line techniques. 
We discuss the prospect of bridging the gap between emission and absorption studies by pushing the emission-based VSF measurements to below $\approx\!10$ kpc.

\end{abstract}



\section{Introduction}
The circumgalactic medium (CGM) is the outermost, gaseous envelope of a galaxy, extending beyond the visible stellar disk and containing the majority of the baryons in the galaxy. This main gas reservoir 
records critical information about a galaxy’s past and ongoing interactions with the surrounding environment. Due to the tenuous nature of the CGM, absorption spectroscopy using bright background sources -- predominantly quasi-stellar objects (QSOs) -- has been the main probe of gaseous halos, yielding sensitive constraints on gas density, temperature, metallicity, and ionization state \citep[e.g.,][]{Chen2017,Tumlinson2017,Rudie2019,Peroux2020,Donahue2022,FaucherGiguere2023}. 

Over the past decade, the advent of wide-field, high-throughput integral field spectrographs (IFSs) has provided a spatial resolving power that complements the pencil-beam probe from QSO absorption spectroscopy, greatly aiding in the investigation of the CGM. Various dynamical processes in the CGM, such as infalls, outflows, and tidal interactions, can now be spatially and spectrally mapped by IFSs via strong nebular emission lines \citep[e.g.,][]{Epinat:2018,Johnson2018,Rupke2019,Chen2019}. One particularly exciting prospect with these resolved kinematic measurements is the robust constraint of turbulent motions in low-density gas.

With a high Reynolds number, ionized, diffuse plasma such as the CGM is expected to be turbulent \citep[see e.g.,][for a recent review]{Burkhart2021}, which can manifest as large density fluctuations commonly observed in extended emission at tens of kpc scales in gaseous halos (e.g., Travascio et al, in prep). Turbulence plays a critical role in several key processes in the CGM, such as mixing/transporting metals \citep[e.g,][]{Pan2010}, facilitating multiphase structure formation \citep[e.g.,][]{Gaspari2018,Fielding2020}, and offsetting radiative cooling \citep[e.g.,][]{Zhuravleva2014}. 
Until recently, observing turbulence in circumgalactic/intergalactic gas has had to rely on two approaches employing high-resolution absorption line spectra of background QSOs.  One approach is to observe line widths of ions with different masses and isolate the turbulent contribution to the velocity profile along the line of sight \citep[e.g.,][]{Rauch:1996, Rudie2019,Qu:2022,Chen:2023b}. Alternatively, if multiple lines of sight (e.g., to gravitationally lensed QSO images) are available, turbulence can be measured as a function of transverse separation between the lines of sight to form the structure functions for the line-of-sight velocities \citep[][]{Rauch2001}. With the advent of IFS, spatially-resolved velocity maps of entire gaseous galactic halos can now be obtained in one shot, enabling the simultaneous measurement of the turbulent power spectrum over a wide range of scales, thus providing multiple independent constraints on the nature of turbulence and the turbulent energy transfer in the gas.

\begin{deluxetable*}{lccrccl}
         \tablenum{1}
        \tablecaption{Summary of the QSO properties. }  
	\label{tab:QSO_summary}
        \tablewidth{0pt}
        \tablehead{
        \colhead{Field name$^a$} & \colhead{$z_{\rm QSO}$} & \colhead{$L_{\rm bol}$ (erg s$^{-1}$)} & \colhead{$N_{\rm group}^b$} & \colhead{$\sigma_{v, {\rm group}}^c$ (km s$^{-1}$)} & \colhead{Radio mode} & \colhead{References} }
        \startdata
		PKS0454$-$22$^*$ & 0.5335 & $\approx  10^{47.0}$ & 23 & $\approx 320$ & Loud & \cite{Helton2021} \\ 
            PKS0405$-$123 & 0.5731 & $\approx  10^{47.3}$ & 20 & $\approx 430$ & Loud & \cite{Johnson2018}\\
            HE0238$-$1904 & 0.6282 & $\approx  10^{47.2}$ & 34 & $\approx 400$ & Quiet & \cite{Liu2023}\\ 
            PKS0552$-$640 & 0.6824 & $\approx  10^{47.4}$ & 10 & $\approx 335$ & Loud & Johnson et al. (2023)\\
            J0454$-$6116$^*$ & 0.7861 & $\approx  10^{46.9}$  & 19 & $\approx 300$ & Quiet & Li et al. in prep. \\ 
		J2135$-$5316$^*$ & 0.8115 & $\approx  10^{47.3}$ & 3 & -- & Quiet & Li et al. in prep. \\  
		TXS0206$-$048$^*$ & 1.1317 & $\approx  10^{47.3}$ & 27 & $\approx 550$ & Loud & \cite{Johnson2022} \\  
        \enddata 
        \tablecomments{\\
        $^a$ VSF analyses for fields marked with $^*$ (i.e., PKS0454$-$22, J0454$-$6116, J2135$-$5316 and TXS0206$-$048) were presented in \citetalias{Chen2023}; PKS0405$-$123, HE0238$-$1904 and PKS0552$-$640 are three newly included fields in this work (see Section \ref{sec: observations_measurements}).\\ 
        $^b$ Number of spectroscopically-identified group member galaxies, including the QSO host.\\
        $^c$ Velocity dispersion of the group.}
\end{deluxetable*}

Recently, \cite{Chen2023} (hereafter Paper I) obtained two-dimensional line-of-sight velocity maps of line-emitting gas around four QSOs up to scales of $\sim 100$ kpc using IFS data. 
Taking advantage of the spatially-resolved velocity maps from IFS observations, these authors constructed 
velocity structure functions (VSFs), $S_p$, defined as 
\begin{equation}
    S_p(r)=\langle | \vect{v}(\vect{x}) - \vect{v}(\vect{x}+\vect{r}) |^p \rangle, 
	\label{eq:vsf}
\end{equation}
where $\vect{x}$ and $\vect{r}$ represent, respectively, a position in the velocity map and the distance vector between two positions separated by $\vect{r}$. 
The exponent $p$ is generally referred to as the order of the VSFs, and $\langle\rangle$ denotes the mean value averaged over all available velocity pairs separated by $r$. As can be seen from the definition, $S_p$ quantifies the scale-dependent variance of a velocity field \citep[e.g.,][]{Frisch1995}, and has been commonly used to probe the dynamical state of the interstellar medium (ISM) in local \ion{H}{2} regions \citep[e.g., ][]{Wen1993, Ossenkopf2002,Federrath2013,Arthur2016,Padoan2016,Melnick2021,Garcia-Vazquez2023} as well as in the intracluster medium (ICM) in nearby cool-core clusters \citep[][]{Li2020,Ganguly2023}. 

While the uncertainties remained large for three QSO nebulae, \citetalias{Chen2023} found that in one particular nebula, the gas dynamics can be unambiguously characterized by the Kolmogorov relation, expected for subsonic, isotropic and homogeneous turbulent flows. 
Building upon the sample studied in \citetalias{Chen2023}, in this follow-up paper, we include results from four nebulae discovered in three new QSO fields. Combining this new sample with the previous one establishes a sample of eight QSO nebulae that allows us to carry out an ensemble study of the empirical properties of CGM turbulence in distant QSO host halos.  
The QSOs are all luminous, with a bolometric luminosity of $\sim 10^{47}$ erg s$^{-1}$, and span a range in redshift from $z\approx 0.5$ to $z\approx 1.1$.  The nebulae are revealed in \oii and/or \oiii line emission (see Figure \ref{fig:nb}) and are selected to have an extended, contiguous emission area $\gtrsim 1500$ kpc$^2$. Table \ref{tab:QSO_summary} summarises the properties of the QSOs in the sample. Out of the seven QSOs, four are radio-loud, and three are radio-quiet.  

This paper is organized as follows. In Section \ref{sec: observations_measurements}, we describe the observations of the ensemble sample and the subsequent velocity measurements using the emission line features.  Based on the spatially-resolved velocity maps, we present the VSFs for all eight nebulae in Section \ref{sec:results}.  We discuss the implications of the results in Section \ref{sec:discussion} and conclude in Section \ref{sec:conclusion}.  Throughout this paper, we adopt a flat $\Lambda$CDM cosmology with 
$H_0=70\rm~ km ~s^{-1}~ Mpc^{-1}$, $\Omega_\mathrm{M}=0.3$ and $\Omega_\Lambda = 0.7$ when deriving distances, masses and luminosities.  All distances quoted are in physical/proper units. 

\section{Observations and data analysis}
\label{sec: observations_measurements}

To constrain the turbulent energy spectrum, we follow the approach described in \citetalias{Chen2023} to construct the VSFs of four nebulae found in three new QSO fields, PKS\,0405$-$123, HE\,0238$-$1904, and PKS\,0552$-$640.  In this section, we briefly summarize the IFS observations and the steps we took to construct a spatially-resolved velocity map based on a line profile analysis of [\ion{O}{2}]$\lambda\lambda$3727, 3729 and [\ion{O}{3}]$\lambda$5008 emission lines in these QSO fields.

\subsection{IFS observations}
To measure the spatially-resolved kinematics in the plane of the sky for the QSO nebulae in our sample, we use the IFS observations obtained using the Multi-Unit Spectroscopic Explorer \citep[MUSE;][]{Bacon2010} on the VLT UT4.  The Wide-Field-Mode (WFM) was used to observe all seven fields, offering a field-of-view (FOV) of $1\arcmin\times1\arcmin$ for a single pointing and a spatial sampling of $0\farcs2$ per pixel.  MUSE covers a wavelength range of 4750--9350\ang\, and has a spectral resolving power of $R\approx 2000$--4000, with a higher resolution at longer wavelengths.

\begin{deluxetable}{lccrc}
	\tablenum{2}
        \tablecaption{Journal of MUSE observations.}
	\tablewidth{0pt}
        \label{tab:QSO_obs_info}
        \tablehead{
        \colhead{} & \colhead{} & \colhead{} & \colhead{} & \colhead{Seeing$^a$} \\
        \colhead{Field name} & \colhead{RA(J2000)} & \colhead{Dec.(J2000)} & \colhead{$t_{\rm exp}$ (s)} & \colhead{(arcsec)}} 
            \startdata
            PKS0454$-$22 & 04:56:08.90 & $-$21:59:09.1 & 2700 & 0\farcs6 \\
            PKS0405$-$123 & 04:07:48.48 & $-$12:11:36.1 & 35100& 0\farcs7\\
            HE0238$-$1904 & 02:40:32.58 & $-$18:51:51.4 & 31500& 0\farcs8\\
            PKS0552$-$640 & 05:52:24.60 & $-$64:02:10.9 & 6000& 0\farcs8\\
		J0454$-$6116 & 04:54:15.95 & $-$61:16:26.6 & 5100 & 0\farcs7 \\
		J2135$-$5316 & 21:35:53.20 & $-$53:16:55.8 & 6840 & 0\farcs6 \\
		TXS0206$-$048 & 02:09:30.74 & $-$04:38:26.5 & 28800 & 0\farcs7\\
		\enddata
    \tablecomments{\\
    $^a$ Atmospheric seeing FWHM measured using the QSO at 7000\ang.  To improve the quality of line fitting, each combined data cube was convolved with a Gaussian kernel of FWHM$=0\farcs7$. This yielded a total PSF FWHM of $\approx 0\farcs9$--$1\farcs0$, corresponding to a projected separation of 6-9 kpc at the redshifts of these QSOs.}
\end{deluxetable}

Table \ref{tab:QSO_obs_info} lists the coordinates, exposure time, and atmospheric seeing conditions of our sample.  Out of the seven QSO fields, the measurements for four fields (PKS0454$-$22, J0454$-$6116, J2135$-$5316, and TXS0206$-$048) were presented in \citetalias{Chen2023}.  The three newly included fields (PKS0405$-$123, HE0238$-$1904, and PKS0552$-$640) are all part of the MUSE Quasar-field Blind Emitters Survey (MUSEQuBES) program, and we use the MUSE-DEEP datacubes directly downloaded from the ESO phase-3 archive with program IDs 097.A-0089(A) and 094.A-0131(B) \citep[PI: J. Schaye;][]{Muzahid2020}.

\subsection{Construction of velocity maps}
\label{sec:velocity_measurements}
As described in 
\citetalias{Chen2023}, the main steps to construct a two-dimensional velocity map include removing the contamination from the QSO point spread function (PSF), subtracting continuum flux across the whole MUSE FOV, constructing optimally extracted narrow-band images for \oii and \oiii lines using three-dimensional masks, and finally fitting Gaussian components to the emission signals and optimizing the parameters via an MCMC analysis.  Readers can find the detailed descriptions and associated technical considerations of each step in \citetalias{Chen2023}. Note that to increase the signal-to-noise ratio for faint spaxels in the outskirts of a nebula, we smooth the data cubes in the spatial dimension with a two-dimensional Gaussian kernel of full-width-at-half-maximum of ${\rm FWHM}=0\farcs7$, leading to a total PSF FWHM of $\approx 0\farcs9$--$1\farcs0$ (see Table \ref{tab:QSO_obs_info}), corresponding to $\approx 6$--9 kpc at the QSO redshifts. 

A subset ($\approx$10--20\%) of spaxels in the nebulae (mostly towards the inner region in the vicinity of the QSOs) exhibit multiple velocity components, which can be identified clearly with the \oiii line.  With MUSE spectral resolution and due to the doublet nature of the \oii line, 
multiple velocity components are only obvious for narrow features with a velocity dispersion $\lesssim 50$ km/s.  In \citetalias{Chen2023}, we demonstrated that different ways of handling the multi-component spaxels (e.g., adopting the flux-weighted mean velocity versus using the velocity of the strongest component) do not lead to significant differences in the VSF measurements.  The insensitivity of the VSFs to the treatment of multi-component spaxels can be attributed to the relatively small proportion of spaxels requiring a multi-component fit, and that the majority of such spaxels exhibit a single prominent component that dominates the kinematics.  Therefore, we opt to take the simple approach of using a single Gaussian function when fitting the lines.  

We also treat \oii and \oiii from the same spaxels separately when conducting the line fitting, allowing the two lines to have different velocities and line widths. This decision is motivated by the observation that for spaxels requiring multiple velocity components, there exists spatial variation in the [\ion{O}{3}]/[\ion{O}{2}] ratio across different components, resulting in a different flux-weighted mean velocity for the two lines.  In addition, the two lines have different footprints within the same nebula due to different signal-to-noise ratios and emission strengths. Therefore, to keep the analyses simple without sacrificing the accuracy of the velocity measurements, we opt to measure [\ion{O}{2}] and [\ion{O}{3}] separately. 

\subsection{VSF measurements}
\label{sec:vsf_measurements}
\begin{figure*}
	\includegraphics[width=0.95\textwidth]{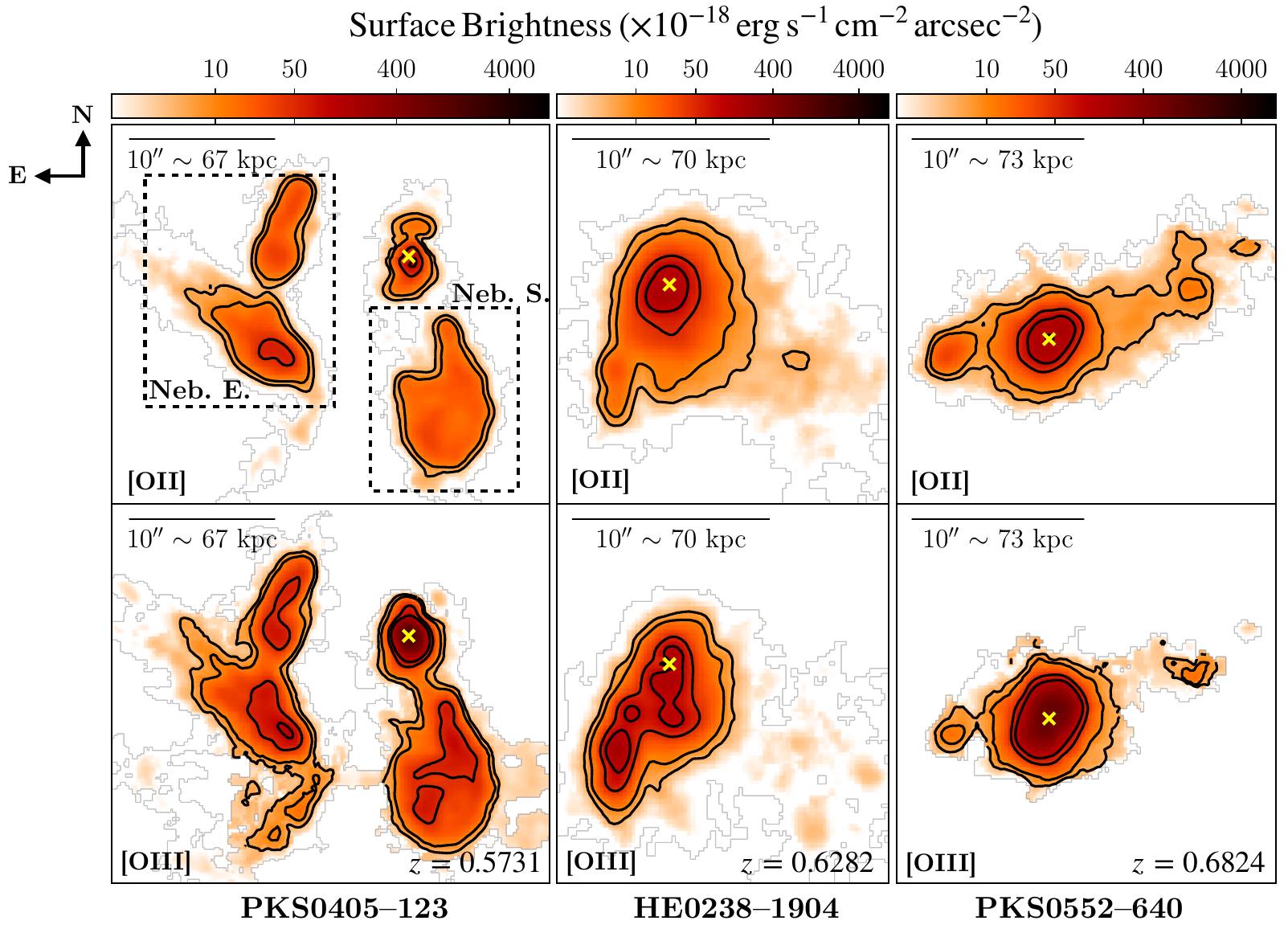}
    \caption{
    Continuum- and QSO-subtracted narrow-band images of the [\ion{O}{2}] and [\ion{O}{3}] emission from the three fields studied in this paper, based on the MUSE-WFM observations. The fields are shown in order of increasing redshift from left to right. Contours are at surface brightness levels of $[5, 10, 50, 100]\times 10^{-18}\,{\rm erg\,s^{-1}\, cm^{-2}\,arcsec^{-2}}$. The yellow cross in each panel marks the quasar position. For PKS0405$-$123, we indicate the eastern nebula (Neb. E.) and the southern nebula (Neb. S.) with dashed boxes (same for both [\ion{O}{2}] and [\ion{O}{3}] emission), as these two nebulae are treated as separate systems despite originating from the same QSO field (see text for details).  The narrow-band images for PKS0454$-$22, J0454$-$6116, J2135$-$5316 and TXS0206$-$048 were presented in \protect\citetalias{Chen2023}. 
    }
    \label{fig:nb}
\end{figure*}

For the three new QSO fields presented in this paper, we show the continuum- and QSO-subtracted narrow-band images in Figure \ref{fig:nb}. The narrow-band images for PKS0454$-$22, J0454$-$6116, J2135$-$5316 and TXS0206$-$048 have already been presented in Figure 1 of \citetalias{Chen2023}. 

As described in Section 3.5 of \citetalias{Chen2023}, to ensure the robustness of the VSF measurements, we exclude spaxels with a velocity uncertainty larger than 45 km/s.  We also examine the velocity map for each nebula in tandem with the broadband images from either MUSE or {\it HST} to identify spaxels that are likely to originate from continuum sources.  If such spaxels exhibit distinctly different velocities and line widths from the rest of the nebula, we exclude them because such continuum sources are likely to 
be separate from the rest of the nebula, and are simply projected to be within the nebula footprint. Finally, we exclude spaxels that are outliers ($\approx 2$ per cent tail on both the blue and red ends) in the probability density distribution of the velocities in each field.  After the above-mentioned steps, all spaxels left in the velocity maps are included in the subsequent VSF calculation, as shown in the top left panels of Figures \ref{fig:PKS0405_seg1_o2}--\ref{fig:PKS0552_o3}.  Summing over all spaxels included in the VSF analyses, the total luminosity and area for each nebula are listed in Table \ref{tab:emission_line_properties}.  

\begin{deluxetable}{lcccc}
        \tablenum{3}
	\tablecaption{Summary of emission properties in QSO nebulae$^a$.}
        \tablewidth{0pt}
        \label{tab:emission_line_properties}
        \tablehead{
        \colhead{} & \multicolumn{2}{c}{Luminosity (erg s$^{-1}$)} & \multicolumn{2}{c}{Nebula area (kpc$^2$)}\\
		 \cline{2-3}\cline{4-5} 
		\colhead{Field name} &\colhead{[\ion{O}{2}]} & \colhead{[\ion{O}{3}]} & \colhead{[\ion{O}{2}]} & \colhead{[\ion{O}{3}]} } 
            \startdata
		PKS0454$-$22 & $1.9\times 10^{42}$ & $2.2\times 10^{43}$ &  1552 & 2202 \\
         PKS0405$-$123 S & $1.2\times 10^{42}$ & $2.8\times 10^{42}$ &  2765 & 3171 \\
         PKS0405$-$123 E & $1.6\times 10^{42}$ & $3.2\times 10^{42}$ &  3839 & 4667 \\
         HE0238$-$1904 & $3.2\times 10^{42}$ & $4.2\times 10^{42}$ &  5081 & 5356 \\
         PKS0552$-$640 & $4.0\times 10^{42}$ & $1.2\times 10^{43}$ &  4105 & 3533 \\
		J0454$-$6116 & $3.5\times 10^{42}$ & $5.3\times 10^{42}$ & 3821 & 2128 \\
		J2135$-$5316 & $2.5\times 10^{42}$ & $9.2\times 10^{42}$ & 1614 & 2190 \\
		TXS0206$-$048  & $2.0\times 10^{43}$ & -- & 6239 & -- \\
		\enddata
	\footnotesize
    \tablecomments{\\
    $^a$ Luminosities and nebula sizes are summed over the spaxels used for the subsequent VSF analyses, which encompass a smaller area than shown in Figure \ref{fig:nb}. Refer to velocity maps (e.g. Figures \ref{fig:PKS0405_seg1_o2}--\ref{fig:PKS0552_o3}) for the regions included in the VSF calculation. }
\end{deluxetable}

Within the spectral coverage of MUSE, we observe both \oii and \oiii emission for six out of seven QSO fields in our sample, and we present the results based on both lines for these fields.  For TXS0206$-$048 at $z\approx 1.1$, the \oiii line is redshifted out of the MUSE spectral window, and therefore only results based on \oii are presented.  PKS0405$-$123 consists of three main nebulae that are cleanly separated in velocity-position space \citep[see Figures \ref{fig:nb} and \ref{fig:PKS0405_seg1_o2}--\ref{fig:PKS0405_seg2_o3}; also see Figure 2 of][]{Johnson2018}. For the purpose of this paper, we analyze the southern and eastern nebulae of PKS0405$-$123 separately and refer to them as PKS0405$-$123 S and PKS0405$-$123 E, and we do not include the nebula immediately surrounding the QSO in this field due to its relatively small size. 

We measure the VSFs up to order $p=6$ for all eight nebulae following the definition of Equation \ref{eq:vsf}. VSFs with $p>6$ become too noisy to provide 
meaningful constraints.  Due to the spatial correlation between spaxels that are separated by distances less than the size of the total PSF, not all velocity pairs in each distance bin are independent.  Therefore, to obtain a more robust estimate of the uncertainty in the VSF measurements, we adopt the modified bootstrap method described in \citetalias{Chen2023}.
In addition, as shown in \citetalias{Chen2023}, the spatial correlation due to atmospheric seeing and the additional Gaussian smoothing applied to the datacubes preferentially removes power from small scales and steepens the VSFs. This smoothing effect can be explicitly accounted for by employing a Gaussian-convolved 
2nd-order VSF, $S_2^\prime$, 
\begin{equation}
    S^\prime_2(r) = 2[\Gamma^\prime(0) - \Gamma^\prime(r)],
\end{equation}
where $\Gamma^\prime$ is a Gaussian-convolved velocity autocorrelation function, 
\begin{equation}
\label{eq:gamma_prime}
    \Gamma^\prime(r)=\Gamma(r)\ast\Gamma_g(r).
\end{equation}
Here $\Gamma(r)$ and $\Gamma_g(r)$ are the autocorrelation function of the velocity field and the smoothing kernel, respectively. A more detailed derivation for Equation \ref{eq:gamma_prime} can be found in Equations 2--7 of \citetalias{Chen2023}.

To quantify the slopes of the 2nd-order VSFs, we adopt a single power-law model:
\begin{equation}
\label{eq:power-law}
    S_2\propto r^{\gamma_2}. 
\end{equation}
When fitting the observed $S_2^\prime$ with a power-law model, we conduct the convolution in Equation \ref{eq:gamma_prime} numerically, and find the best-fitting $\gamma_2$ with the \texttt{Scipy curve\_fit} routine for each of the 1000 modified bootstrap samples described above to obtain the mean and dispersion of $\gamma_2$.  Note that we only consider non-negative slopes of $\gamma_2$, which is motivated by data and avoids divergent values at $r=0$. 

With the IFS data, observations are confined to projected quantities both in velocity and spatial separations. Therefore, we report the VSF measurements using the line-of-sight velocities and the projected spatial separation $r_{\rm proj}$ in the plane of the sky. The potential limitations due to the projection effect will be discussed in further detail in Section \ref{sec:limitation_caveats}. 

\label{sec:results_vsf}
\begin{figure*}
	\includegraphics[width=\textwidth]{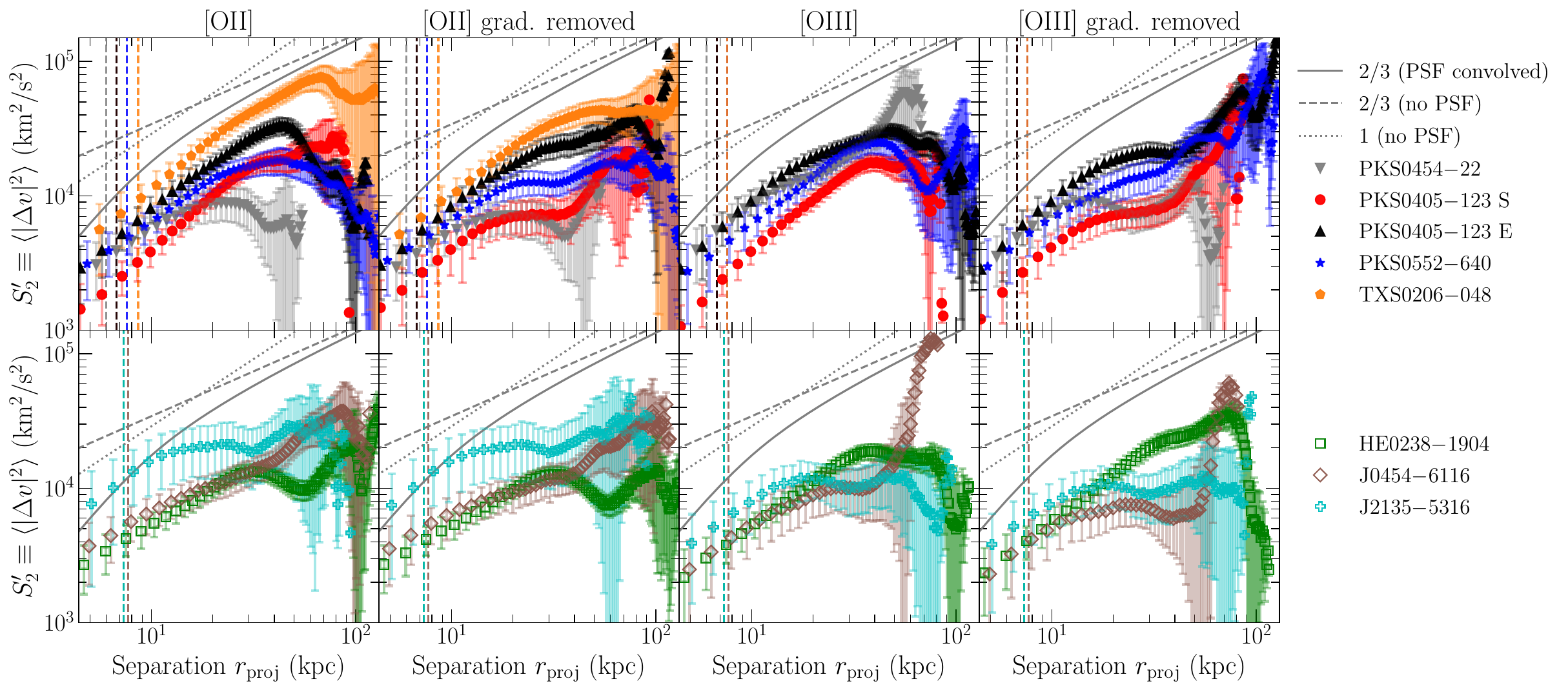}
    \caption{{\it Top row: }The observed 2nd-order VSF $S_2^\prime(r)$ for the nebulae of radio loud QSOs.  Vertical dashed lines mark the size of the total PSF FWHM for each field.  The data points and the error bars show the mean and the standard deviation for the 1000 measurements obtained through the modified bootstrap method (see Section \ref{sec: observations_measurements}). The dashed and solid gray lines show, respectively, the expected $S_2$ for Kolmogorov turbulence before and after convolving with an appropriate PSF.  As different fields have slightly different PSF sizes, we use the mean value of the PSF FWHM for all radio-loud fields included in the panels when constructing the expected Kolmogorov VSF.  The dotted gray lines show a power-law with a slope of 1 (i.e., Burger's turbulence).  The four panels from left to right show the results using the [\ion{O}{2}] and [\ion{O}{3}] lines, both from direct measurements and after removing a uni-directional gradient (see text). {\it Bottom row: }Same as the top row but for radio-quiet fields. All nebulae exhibit VSFs with an increasing amplitude in velocity variance as a function of separation distance. PSF smoothing significantly steepens the apparent slopes of the VSFs and we will explicitly take into account this smoothing effect when fitting the VSFs with a power-law (see Section \ref{sec:s2_slopes}).}
    \label{fig:S2}
\end{figure*}

\section{Results}
\label{sec:results}
Using the velocity maps constructed for individual nebulae, we proceed with the VSF analysis using the full sample of eight extended nebulae. Recall that while it is relatively straightforward to measure the VSFs using spatially-resolved velocity maps, a primary systematic uncertainty is possible contributions to the observed signal from coherent bulk motions projected in the plane of the sky. To account for this uncertainty, we follow the approach adopted by \citetalias{Chen2023} and consider a simple, unidirectional velocity gradient model parameterized as $v(x,y)=ax+by+c$, where $x$ and $y$ are coordinates of individual spaxels, and $a$, $b$, and $c$ are the free parameters.  For each nebula, we measure the VSFs with and without the best-fitting two-dimensional bulk-flow model subtracted. 
The amplitudes of the best-fitting gradient for the \oii and \oiii emission lines in each field are listed in Table \ref{tab:slopes}. We estimate the uncertainty of this velocity gradient by repeating the fitting with 1000 randomly perturbed velocity maps based on the MCMC chains for each spaxel and find that the uncertainties are small ($\lesssim$ 0.1 km s$^{-1}$) for all nebulae.  Therefore, we do not list the uncertainties in Table \ref{tab:slopes}.
To identify possible coherent motions dominant along the radial or tangential directions (for example in the case of strong outflows or inflows), \citetalias{Chen2023} also calculated the VSFs using radial and tangential velocity pairs separately and found the VSF measurements to be comparable along these two directions.  For the newly analyzed nebulae in this paper, we find a similar trend where radial and tangential VSFs show no clear differences and are therefore not included in the presentation here.  

In this section, we first examine the general trend displayed in the second-order VSF across all eight QSO nebulae.  Then we quantify and compare the best-fitting VSF slope over a finite range of spatial scale where the measurements can be characterized by a power-law function.  Finally, we explore the presence or absence of extended self-similarity \citep[ESS; see, e.g.,][]{Benzi1993} in turbulent flows in QSO host halos by measuring the higher-order VSFs.

\subsection{The overall shape of VSFs}

Figure \ref{fig:S2} shows the observed 2nd-order VSFs, $S^\prime_2$, for the eight nebulae in our sample. Radio-loud and radio-quiet fields are shown in the top and bottom rows, respectively. The vertical dashed lines mark the FWHM of the total PSF for each field (see Table \ref{tab:QSO_obs_info}).  To guide the visual comparison, we overplot the expected $S_2$ for Kolmogorov turbulence, with the dashed gray line showing the intrinsic 2/3 slope and the solid gray line showing the observed shape of $S_2^\prime$ after convolving with an appropriate PSF.  Because different fields have slightly different PSF sizes, we use the mean value of the PSF FWHM for radio-loud (-quiet) fields when constructing the expected Kolmogorov $S_2^\prime$ for the top (bottom) row.  We also show the power-law with a slope of 1 (e.g., Burger's turbulence), without convolving with a PSF, in dotted gray lines.  The comparison between the data and the model $S_2$ with slopes 2/3 and 1 underlines the importance of including the PSF effect when quantifying the observed VSF slopes.  In particular, if the probed distance separation, $r_{\rm proj}$, is $\lesssim 10$--20 times the PSF FWHM, the PSF smoothing effect can significantly steepen the apparent slope of the VSFs and a naive visual inspection will lead to the wrong conclusion that the VSF slopes are steeper than their intrinsic values. The VSFs obtained using the gradient-removed velocity maps are also included in Figure \ref{fig:S2} for comparison. 

As shown in Figure \ref{fig:S2}, all nebulae in our sample exhibit an overall increasing trend of velocity fluctuations with increasing spatial scale.  The values of $\langle \Delta v^2\rangle$ range from $\approx 5000$--10,000 km$^2$/s$^2$ at $r_{\rm proj}\approx10$ kpc to $\approx 10,000$--80,000 km$^2$/s$^2$ at $r_{\rm proj}\approx50$ kpc.  The results based on the \oii and \oiii lines are consistent within the uncertainty for fields with both lines. In general, we do not expect the VSFs constructed from \oii and \oiii lines to be identical, because the footprints of the two emission lines in the nebulae do not overlap completely due to the different signal-to-noise ratios of the two lines at different locations.  For regions with overlapping footprints from both \oii and \oiii emission, the line-of-sight velocities can also differ for spaxels with multiple velocity components and varying [\ion{O}{3}]/[\ion{O}{2}] line ratios between components, as discussed in Section \ref{sec:velocity_measurements}.  We will show below that the VSFs from [\ion{O}{2}] and [\ion{O}{3}] lines lead to consistent constraints on the dynamical state of the gas.  In addition, the removal of a large-scale, unidirectional velocity gradient generally flattens the VSFs via preferentially reducing the power at larger distance separations.  Nonetheless, the constrained slopes for a single power-law fit are consistent before and after the removal of the gradient, as we will discuss in the following section. 

\begin{deluxetable*}{lcccccccccc}
        \tablenum{4}
	\tablecaption{Summary of the power-law slopes of the VSFs constructed using [\ion{O}{2}] and [\ion{O}{3}] lines$^a$. }  
        \tablewidth{0pt}
	\label{tab:slopes}
        \tablehead{
		 & \multicolumn{3}{c}{[\ion{O}{2}]} & \multicolumn{2}{c}{[\ion{O}{2}] grad.\ removed$^b$} & \multicolumn{3}{c}{[\ion{O}{3}]} & \multicolumn{2}{c}{[\ion{O}{3}] grad.\ removed} \\
         \cmidrule(lr){2-4}\cmidrule(lr){5-6}\cmidrule(lr){7-9}\cmidrule(lr){10-11}
		Field name &[$r_1$, $r_2$]$^c$ & $\gamma_2$ & Gradient$^d$ & [$r_1$, $r_2$] &$\gamma_2$ &[$r_1$, $r_2$] &$\gamma_2$ & Gradient &[$r_1$, $r_2$] &  $\gamma_2$}
        \startdata
        PKS0454$-$22       & [5.8, 20] & $<0.78$  & 2.2 & [5.8, 17] & $<0.66$ & [5.8, 20] & $<0.67$ & 5.0 & [5.8, 14] & $<1.45$ \\[0.15cm]
        PKS0405$-$123 S & [7.4, 29] & $1.07^{+0.20}_{-0.18}$ & 5.8 & [7.4, 17] & $<1.54$ & [7.4, 34] & $0.97^{+0.15}_{-0.15}$ & 6.2 & [7.4, 17] & $<1.41$ \\[0.15cm]
        PKS0405$-$123 E & [7.4, 37] & $0.76^{+0.19}_{-0.16}$ & 6.0 & [7.4, 30] &$0.55^{+0.22}_{-0.21}$ & [7.4, 46] & $0.33^{+0.11}_{-0.11}$ & 5.8 & [7.4, 22] & $<1.04$ \\[0.15cm]
        HE0238$-$1904 & [8, 29] & $0.48^{+0.17}_{-0.18}$ & 0.9 & [8, 30] &$0.43^{+0.18}_{-0.18}$ & [8, 33] & $0.75^{+0.15}_{-0.15}$ & 2.2 & [8, 33] & $0.88^{+0.17}_{-0.17}$ \\[0.15cm]
        PKS0552$-$640 & [8.3, 25] & $0.55^{+0.28}_{-0.28}$ & 5.0 & [8.3, 22] &$<0.97$ & [8.3, 32] & $0.88^{+0.20}_{-0.22}$ & 8.7 & [8.3, 37] & $<0.50$ \\[0.15cm]
        J0454$-$6116 & [7.5, 30] & $<0.51$ & 1.6 & [7.5, 40] & $<0.45$ & [7.5, 25] & $<0.84$ & 2.8 & [7.5, 25] & $<0.33$\\[0.15cm]
        J2135$-$5316 & [7.2, 25] & $<0.50$ & 0.9 & [7.2, 23] &  $<0.65$ & [7.2, 18] & $<1.23$ & 1.8 & [7.2, 18] & $<1.12$\\[0.15cm]
        TXS0206$-$048 & [8.5, 60] & $0.72^{+0.12}_{-0.11}$ & 3.7 & [8.5, 40] & $0.56^{+0.16}_{-0.17}$ & -- &--&--&--&--  \\[0.15cm]
	\enddata
    \tablecomments{\\
    $^a$ The best-fitting slopes are derived from 1000 modified bootstrap samples, as discussed in Section \ref{sec: observations_measurements}. These slopes correspond to the intrinsic power-law slopes for $S_2$, with our fitting process explicitly addressing the PSF smoothing effect in the measured $S^{\prime}_2$. The reported values are medians along with the 16$^{\rm th}$ and 84$^{\rm th}$ percentiles.  The 3$^{\rm th}$ and $97^{\rm th}$ percentiles are approximately double the uncertainty estimates listed here for all fields. For the unconstrained results, we present 95\% upper limits for the slope, assuming the observed pair separations fall within the inertial range.  If the available pair separations are close to injection scales, then no robust constraints can be obtained.  We exclusively consider non-negative power-law slopes, in line with the discussion in Section \ref{sec: observations_measurements}.\\
    $^b$ Measurements obtained after removing a 2D velocity gradient (see \S\,\ref{sec:results}).\\
    $^c$ Lower and upper bounds in the projected distance separation, $r_{\rm proj}$, in the unit of kpc, within which the power-law slopes of the VSFs are constrained (see \S\,\ref{sec:s2_slopes}).\\  
    $^d$ Best-fitting 2D velocity gradient, in the unit of km/s/kpc. } 
\end{deluxetable*}

\subsection{2nd-order VSF slopes}
\label{sec:s2_slopes}
As shown in Figure \ref{fig:S2}, all VSFs exhibit structures that deviate from a single power-law.  In particular, at larger separations, the VSFs can show an overall decreased amplitude (e.g., TXS0206$-$048 at $r_{\rm proj}\gtrsim 60$ kpc), an overall enhanced power (e.g., J0454$-$6116 at $r_{\rm proj}\gtrsim 30$ kpc), or an oscillatory behavior (e.g., HE0238$-$1904 at $r_{\rm proj}\gtrsim 30$ kpc).  Such deviations may reflect different levels of velocity fluctuations in the central regions of the nebulae versus the outskirts, as velocity pairs at larger separations are predominantly constructed from spaxels in the outskirts.  In addition, large-scale periodic oscillations in the velocity fields can manifest as oscillations in the VSFs at large separations \citep[e.g., ][]{Garcia-Vazquez2023}.  The VSF measurements at larger separations are also more uncertain due to a combined effect of fewer pair counts and uncertain velocity centroids as a result of fainter signals in the outskirts of a nebula.

Taking into account the above-mentioned factors, we restrict the fitting to be within a finite range of spatial scales, [$r_1$, $r_2$], 
when employing a single power-law model to quantify the slopes of the VSFs.  The lower limit $r_1$ is chosen to be the FWHM of the total PSF for each field (see Table \ref{tab:QSO_obs_info}), while the upper limit $r_2$ is chosen through a series of trial and error such that we obtain the lowest reduced $\chi^2$ for the best-fitting model within this range.  We refer to $r_2$ as the VSF turnover scale and will discuss its correlation with the nebula size later in Section \ref{sec:turnover_correlation}.  When constraining the VSF slopes, we explicitly incorporate the smoothing effect in the 2nd-order VSF models before comparing them with the data, as described in Section \ref{sec:vsf_measurements}. The [$r_1$, $r_2$] values as well as the best-fitting slopes for the 2nd-order VSF, $\gamma_2$, are listed in Table \ref{tab:slopes} using both the directly measured line-of-sight velocity maps and the gradient-removed velocity maps. As mentioned above, removing a large-scale, unidirectional gradient tends to flatten the VSF, leading to a smaller $r_2$ and weaker constraints on the VSF slopes. The comparisons between best-fitting power-law models and the data for PKS0454$-$22, J0454$-$6116, J2135$-$5316, and TXS0206$-$048 are shown in \citetalias{Chen2023}, while the models for PKS0405$-$123 S, PKS0405$-$123 E, HE0238$-$1904, and PKS0552$-$640 are shown in Figures \ref{fig:PKS0405_seg1_o2}--\ref{fig:PKS0552_o3} in the Appendix of this paper.  

Based on the line-of-sight velocity maps directly measured using the \oii and \oiii emission lines (top left panels of Figures \ref{fig:PKS0405_seg1_o2}--\ref{fig:PKS0552_o3}), the slope $\gamma_2$ for the eight nebulae in our sample shows a range of values. Specifically, the 16$^{\rm th}$--84$^{\rm th}$ measurement percentiles of four nebulae are consistent with the Kolmogorov expectation of $\gamma_2=2/3$ (PKS\,0405$-$123 E, HE\,0238$-$1904, PKS\,0552$-$640, and TXS\,0206$-$048), while three nebulae show flatter VSFs (PKS\,0454$-$22, J0454$-$6116 and J2135$-$5316).  PKS\,0405$-$123 S exhibits a steeper slope but this is also a system that shows a large-scale velocity gradient across the nebula.  After removing a unidirectional velocity gradient, the VSF is consistent with the Kolmogorov expectation.  Below we discuss these three categories individually.

{\it Nebulae with $\gamma_2$ consistent with 2/3: } the VSF measurements for PKS0405$-$123 E, HE0238$-$1904, PKS0552$-$640 and TXS0206$-$048 lead to a constrained 2nd-order slope in agreement with the value 2/3.  For HE0238$-$1904 and PKS0552$-$640, the measurements for both \oii and \oiii within the 16$^{\rm th}$--$84^{\rm th}$ percentiles are consistent with the Kolmogorov slope.  For TXS0206$-$048, only measurements with \oii are available and the result is consistent with $\gamma_2=2/3$. While the VSF slope for the nebula PKS0405$-$123 E based on \oiii is flatter than 2/3 within the 16$^{\rm th}$--$84^{\rm th}$ percentiles, the values within the 3$^{\rm th}$--$97^{\rm th}$ percentiles using both \oii and \oiii emission are in agreement with the Kolmogorov slope and therefore we consider the VSFs of this nebula consistent with the Kolmogorov expectation. 
  
{\it Nebulae with $\gamma_2<2/3$: }for the three nebulae in PKS\,0454$-$22, J0454$-$6116 and J2135$-$5316, only upper limits of $\gamma_2$ can be obtained and the 95\% limits derived from \oii measurements are below 2/3.  While the $\gamma_2$ upper limits obtained from the \oiii measurements are larger than 2/3, the smaller $\gamma_2$ upper limits obtained using \oii suggest that the VSF slopes for these three nebulae are likely flatter than the Kolmogorov expectation.  As we discussed in \citetalias{Chen2023}, the flatter VSFs may indicate the presence of multiple energy injection scales \citep[e.g,][]{ZuHone2016} and/or the effect of a dynamically important magnetic field \citep[e.g.,][]{Boldyrev2006,Brandenburg2013,Grete2021,Mohapatra2022}. 

\begin{figure*}
	\includegraphics[width=\textwidth]{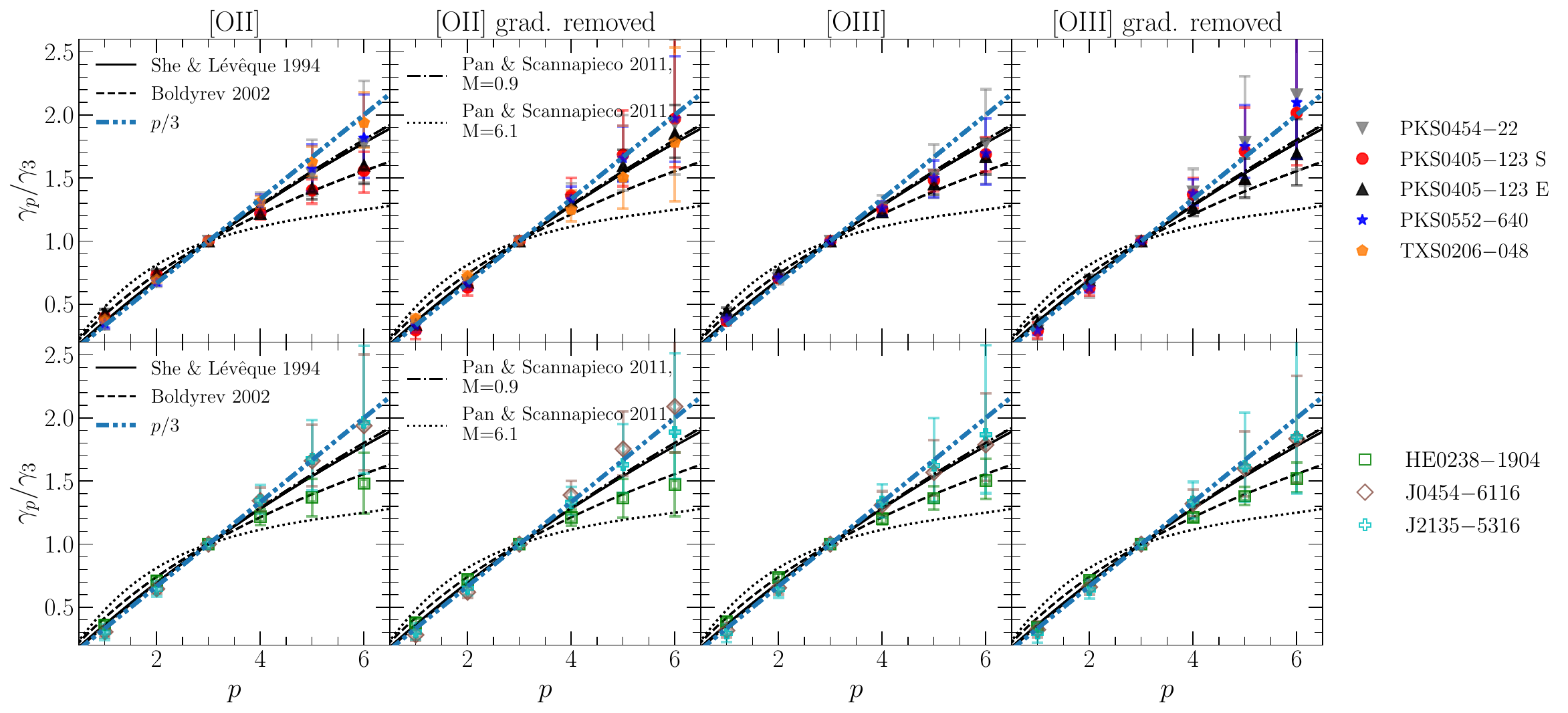}
       \caption{Ratios $\gamma_p/\gamma_3$ for all eight nebulae in the sample based on both \protect\oii and \protect\oiii measurements, as well as their corresponding velocity residual maps after removing a coherent, unidirectional gradient.  $\gamma_p/\gamma_3$ is the best-fitting power-law slope for the relation between the observed $p$th-order VSF, $S_p^\prime$, and $S_3^\prime$.  The data points represent the median values obtained from fitting the 1000 modified bootstrap samples (see Section \ref{sec:vsf_measurements}), and the error bars indicate the 16$^{\rm th}$ and 84$^{\rm th}$ percentiles.  The solid curves represent the expected ratio of $\gamma_p/\gamma_3$ for subsonic Kolmogorov turbulence, taking into account the intermittency correction presented in \protect\cite{SheLeveque1994}.  The dashed curves represent the expected ratio for supersonic magnetohydrodynamic turbulence, as presented in \protect\cite{Boldyrev2002}.  The dash-dotted (dotted) curves indicate the $\gamma_p/\gamma_3$ ratio derived from numerical hydrodynamic turbulent simulations for a Mach number of 0.9 (6.1), as presented in \protect\cite{PanScannapieco2011}.  Finally, the blue dash-dotted curves represent the expected $\gamma_p/\gamma_3$ ratio for Kolmogorov turbulence without the intermittency correction, scaling simply as $p/3$.  The top row shows the results for radio-loud fields, while the measurements for radio-quiet fields are shown at the bottom. Except for the field of HE0238$-$1904, all nebulae exhibit ESS slope ratios in agreement with expectations for subsonic motions using directly measured velocity maps and/or velocity residual maps after removing a coherent unidirectional gradient. None of the nebulae show signatures of supersonic motions with a Mach number $\gtrsim6$.}
    \label{fig:ess}
\end{figure*}

{\it Nebulae with $\gamma_2>2/3$: }Based on the directly measured velocity fields, PKS0405$-$123 S exhibits VSFs that are steeper than the expectation of Kolmogorov turbulence. The constraints are consistent using the \oii and \oiii measurements.  One possible explanation for the steepening of the VSF in this nebula is a strong effect of projection smoothing if the depth of the nebula is larger than the projected distance scales in the plane of the sky (see discussions in Section \ref{sec:limitation_caveats}). Moreover, the line-of-sight velocity maps for both [\ion{O}{2}] and [\ion{O}{3}] show a possible velocity shear along the NW-SE direction (see Figures \ref{fig:PKS0405_seg1_o2} and \ref{fig:PKS0405_seg1_o3}).  The best-fitting direction and amplitude for the velocity gradient are consistent between both emission lines, suggesting that the bulk flow can plausibly contribute to the VSF measurements, leading to steeper VSF slopes.  Indeed, the VSFs become flatter after we remove a unidirectional velocity gradient, resulting in slope upper limits consistent with the Kolmogorov expectation albeit with larger uncertainties. 

In summary, using the direct measurements of the line-of-sight velocity fields based on the \oii and/or \oiii emission lines, five out of eight nebulae exhibit a 2nd-order VSF slope that is consistent with the expected value of 2/3 for Kolmogorov turbulence while three exhibit a flatter VSF.  Incidentally, the three nebulae with a flatter VSF are also the smallest in the sample (see Table \ref{tab:emission_line_properties}).  It is possible that the observations do not have a sufficiently large dynamic range for securing a robust constraint on the shape of the VSF \citep[see, e.g.,][]{Federrath2021}.


\subsection{Extended self-similarity (ESS) in turbulent flows}
\label{sec:ess_slopes}
In addition to measuring the 2nd-order VSF slope $\gamma_2$, \citetalias{Chen2023} also explored the presence of ESS, in which a simple power-law function holds between VSFs of different orders
on spatial scales that are outside of the inertial range where the Kolmogorov relation applies \citep[see, e.g.,][]{Benzi1993}.  This ESS is particularly useful for inferring the energy cascade rate when the inertial range is not well established.
Compared with the slopes of VSFs of individual orders, the ESS slope ratios are often better constrained with a higher statistical significance thanks to the tight correlation between different orders. In addition, an enhanced level of intermittency in a velocity field will suppress the VSF slopes at higher orders compared with the slopes of lower orders \citep[e.g.,][]{Frisch1995}, making the ESS slope ratios a valuable diagnostic for the underlying gas dynamics. Here we explore the presence or absence of ESS in the QSO nebulae by measuring the VSFs up to order $p=6$.  We obtain the slope ratios $\gamma_p/\gamma_3$ for $p=1$--6 by fitting a single power-law model to the $S_p^\prime$ vs. $S_3^\prime$ measurements.  As discussed in \citetalias{Chen2023}, the smoothing effect due to the data PSF does not change the ESS slope ratios. The results are displayed in Figure \ref{fig:ess}, where the data points represent the median values obtained from fitting the 1000 modified bootstrap samples (see Section \ref{sec:vsf_measurements}), and the error bars indicate the 16$^{\rm th}$ and 84$^{\rm th}$ percentiles. 
The correlation between 2nd- and 3rd-order VSFs for each nebula are displayed in the right-most panels of Figures \ref{fig:PKS0405_seg1_o2}--\ref{fig:PKS0552_o3}. 

Specifically, we measure $\gamma_p/\gamma_3$ using the \oii and \oiii velocity maps as well as their corresponding residual maps after removing a unidirectional velocity gradient. Figure \ref{fig:ess} shows the ESS slope ratios, with radio-loud fields in the top row and radio-quiet fields at the bottom.  We also overplot the expected $\gamma_p/\gamma_3$ ratios from different theoretical considerations and numerical simulations, including the Kolmogorov expectation of $\gamma_p/\gamma_3=p/3$ (blue dashed curve), the Kolmogorov turbulence with intermittency correction \citep[solid curve;][]{SheLeveque1994}, the expectation for supersonic magnetohydrodynamic turbulence \citep[dashed curve;][]{Boldyrev2002}, and numerical predictions for hydrodynamic turbulence with Mach numbers of ${\cal M}=0.9$ and 6.1 \citep[dash-dotted and dotted curves;][]{PanScannapieco2011}.  In general, the ratio $\gamma_p/\gamma_3$ is expected to be suppressed significantly at larger $p$'s in supersonic flows with a high Mach number.  This can be seen in Figure \ref{fig:ess} where the numerical simulations predict that for gas motions with ${\cal M}=6.1$, $\gamma_p/\gamma_3$ does not increase significantly for $p>3$, showing a plateau in the $\gamma_p/\gamma_3$ curve (dotted lines). 

While the strongest distinguishing power for different scenarios comes in at higher orders, the measurements are also more uncertain. In addition, removing a large-scale gradient from the velocity field can change the $\gamma_p/\gamma_3$ ratios to be more consistent with predictions for lower Mach numbers (e.g., see the trend for PKS0405$-$123 S). Within the 16$^{\rm th}$ and 84$^{\rm th}$ measurement percentile range and considering the results both before and after removing the large-scale velocity gradient, seven out of eight nebulae in our sample show ESS slope ratios consistent with expectations from subsonic turbulence (black solid curve, blue dash-dotted curve, and dash-dotted curve in Figure \ref{fig:ess}).  For the nebula surrounding HE0238$-$1904, the $\gamma_p/\gamma_3$ ratios are consistent with the predictions for supersonic magnetohydrodynamic turbulence as presented in \cite{Boldyrev2002}, suggesting that the Mach number of gas motions in this field may be higher than that in other nebulae. Given that this field has a constrained $\gamma_2$ value that is consistent with the Kolmogorov expectation as discussed above, additional effects (e.g., the presence of a dynamically important magnetic field) might contribute to a relatively small $\gamma_2$ in tandem with suppressed $\gamma_p/\gamma_3$ ratios. A more detailed investigation into the properties of this nebula (e.g., ionization state, interactions with group member galaxies) is needed to further shed light on the possible physical causes for this difference in ESS slope ratios, and a larger sample is required to examine whether the HE0238$-$1904 nebula is a special case. Overall, no system in our sample exhibits ESS slope ratios that indicate gas motions with ${\cal M}\gtrsim6$.

\section{Discussion}
\label{sec:discussion}

We have shown that the 2nd-order VSF measured for eight QSO nebulae in our sample exhibits a range of slopes.  While five of the nebulae in our sample are consistent with the expected slope of 2/3 for Kolmogorov turbulence, the remaining three exhibit a shallower slope.  Despite a range of 2nd-order VSF observed in these QSO nebulae, the measurements suggest that turbulent flows in the \oii and \oiii line-emitting clouds are subsonic. The subsonic dynamical state of the gas is further corroborated by the ESS slope ratios $\gamma_p/\gamma_3$, which are consistent with theoretical or numerical expectations for subsonic systems with ${\cal M}\lesssim1$ in seven out of eight nebulae. None of the systems shows $\gamma_p/\gamma_3$ measurements that are indicative of highly supersonic flows with ${\cal M}\gtrsim6$. 

In addition, we do not observe significant differences between radio-loud and radio-quiet QSO fields in terms of nebula size, line emission luminosity, VSF slopes, VSF amplitude, and turbulent energy heating rate.  Recall that five of the nebulae in our sample occur near radio-loud QSOs, while the remaining three reside in radio-quiet halos.  The main distinguishing characteristic between radio-loud and radio-quiet QSOs is the presence of powerful jets in radio-loud sources that can result in large-scale structures like radio lobes spanning from tens to thousands of kpc in size \citep[e.g.,][]{Mullin2008}. The mechanical energy contained in the collimated jets and the associated inflated bubbles is estimated to be $\sim 10^{41}$--$10^{46}$ erg/s \citep[e.g., ][]{Heckman2014}. If a significant portion of this energy can be deposited into the CGM as kinematic energy, we may expect the VSFs from radio-loud and radio-quiet fields to exhibit different properties.
While previous studies have found that radio jets are the dominant mechanism for driving fast outflows in the inner $\lesssim10$ kpc regions in radio galaxies \citep[e.g.,][]{Nesvadba2017}, a lack of correlation between the observed VSFs and the radio power suggests that the effect of radio jets may be limited to the inner regions and
have little influence on the gas kinematics on scales $\gtrsim$tens of kpc. This is in agreement with simulation predictions for the ICM in cool-core clusters \citep[e.g.,][]{Yang2016}. A larger sample with both radio-loud and radio-quiet sources will be helpful to draw robust conclusions regarding the difference (or lack thereof) in the CGM dynamics between these two populations. 

In this section, we first discuss the implications for the dynamical state of the gas in the multiphase CGM and infer the energy transfer rate in these QSO host nebulae.  We then discuss potential caveats associated with observational limitations, including projection effects, finite nebula sizes, and the small number of systems in the current sample. 
 
\subsection{Implications for the multiphase CGM dynamics}
\label{sec:discussion_dynamics}
Based on the velocity dispersion of member galaxies in the QSO host group environment, the halo mass of the QSO hosts in our sample is estimated to be $\approx10^{13}$--$10^{14}\msun$ \citep[see e.g., ][]{Johnson2018,Helton2021,Johnson2022,Liu2023}.  This mass range suggests a viral temperature of $T\approx 10^6$--10$^7$ K for the underlying hot halo \citep[e.g.,][]{Mo2010}. Meanwhile, the sound speed of the gas can be calculated by $c_{\rm s}=\sqrt{\gamma k_{\rm B}T/\mu m_{\rm p}}$, where $\gamma=5/3$ is the adiabatic index for an ideal monatomic gas, $k_{\rm B}$ is the Boltzmann constant, $\mu$ is the mean atomic weight (which is 0.588 for fully ionized gas), and $m_{\rm p}$ is the proton mass. For the cool gas of $T\approx10^4$ K, $c_{\rm s, cool}\approx15$ km/s, while for the hot medium of $T\approx 10^6$--10$^7$ K, $c_{\rm s, hot}\approx150$--500 km/s. Therefore, for the nebulae in our sample, the Mach number calculated using the sound speed of the cool gas is $\mathcal{M}_{\rm cool} = \sqrt{3}\sigma_{\rm pos}/c_{\rm s, cool}\approx 7$--18, and $\mathcal{M}_{\rm hot} = \sqrt{3}\sigma_{\rm pos}/c_{\rm s, hot}\approx 0.2$--1.8 using $c_{\rm s, hot}$ for the hot gas. 
Here $\sigma_{\rm pos}$ is the velocity dispersion in the plane of the sky.  As we will discuss below in Section \ref{sec:velocity_dispersion}, $\sigma_{\rm pos}$ is typically smaller than the velocity dispersion along the line of sight, and the Mach numbers will be larger ($\mathcal{M}_{\rm cool}\approx 9$--20 and $\mathcal{M}_{\rm hot}\approx 0.3$--2.0) when estimated using the line-of-sight velocity dispersion.

Given the contrast between the two Mach numbers, $\mathcal{M}_{\rm cool}$ and $\mathcal{M}_{\rm hot}$, the subsonic motions revealed by the VSFs of the nebulae suggest that the [\ion{O}{2}] and [\ion{O}{3}] emission originates from cool gas clumps embedded in the ambient hot medium.  If these cool clumps are in pressure equilibrium with the hot halo, then they can serve as tracers for the kinematics of the volume-filling plasma. The scenario of a dynamically-coupled multiphase gaseous system is supported by absorption line studies on CGM kinematics of $z\sim2$ star-forming galaxies \citep[e.g.,][]{Rudie2019} as well as by recent measurements in the core regions of nearby galaxy groups and clusters \citep[e.g.,][]{Li2020,Olivares2022}.  There has also been an increasing number of theoretical and numerical predictions arguing for a shared dynamical state across different gas phases \citep[e.g.,][]{Gaspari2018,Gronke2018,Schneider2020,Mohapatra2022}

The dynamical coupling likely happens due to a combination of physical processes involving cooling, the exchange of mass and momentum between cool and hot phases, and the competition between cool clump formation and cloud crushing at different mass/length scales. Turbulence is expected to facilitate these processes, which in turn further feed into the development of turbulence in the gaseous halo.  In the absence of turbulence, the condensed cool clumps tend to settle in more organized structures such as a disk. The extended morphological features of the nebulae in our sample suggest that turbulence is significant in these gaseous halos. Phenomenologically, \cite{Gaspari2018} proposed an empirical criterion of $t_{\rm cool}/t_{\rm eddy}\lesssim 1$ for the condensation and survival of cool gas in clusters and groups, where $t_{\rm cool}$ is the gas cooling time and $t_{\rm eddy}$ is the eddy turnover time.  Based on the VSF measurements, we can calculate the eddy turnover time via $t_{\rm eddy} \approx \epsilon^{-1/3}l^{2/3}$, where $\epsilon$ is the energy transfer rate per unit mass at the spatial scale $l$ (for more discussion on $\epsilon$ see Section \ref{sec:epsilon} below). For the nebulae in our sample, we estimate $t_{\rm eddy}\approx 60$--150 Myr at $l\approx 10$ kpc and $t_{\rm eddy}\approx 150$--300 Myr at $l\approx 50$ kpc. While we cannot obtain an estimation for $t_{\rm cool}$ due to the absence of temperature and metallicity measurements of the hot phase, our measured $t_{\rm eddy}$ is in agreement with the estimated values ($t_{\rm eddy}\approx100$--200 Myr for galaxy groups) that fulfill the gas condensation criterion in \cite{Gaspari2018} (see their Figure 5). 


In addition, turbulence in the CGM can also be produced by Kelvin-Helmholtz instability during the accretion of cool gas streams \citep[e.g.,][]{Vossberg2019,Mandelker2019,Li2023}, and motions of fragmented cool gas clumps in disrupted, turbulent mixing zones near the accreting streams are predicted to be subsonic in numerical simulations \citep[e.g.,][]{Aung2019}.  Among our sample, the nebula in the field of TXS0206$-$048 exhibits compelling signs of cool, filamentary gas accretion from large scales \citep[][]{Johnson2022}, suggesting that the observed subsonic turbulence may be in part produced through the accreting streams. 

Finally, previous studies have identified a correlation between the presence of close companions around the QSOs and the presence of strong, extended nebular line emission \citep[see e.g., a narrow-band imaging survey by][]{Stockton1987}. In our sample, the morphokinematics of some nebulae (e.g., PKS0405$-$123, HE0238$-$1904, TXS0206$-$048) reveal that part of the line-emitting gas originates from stripped ISM of group member galaxies as indicated by consistent line-of-sight velocities between the galaxies and extended nebulae \citep[see e.g.,][]{Johnson2018,Helton2021,Liu2023}. It is natural to assume in these cases that the tidal interactions between group member galaxies disturb the gas and enhance turbulence and thermal instabilities in the hot halo, leading to more efficient cooling and cool clump condensation.  The stripped ISM can also serve as massive cool gas seeds that facilitate the coagulation of smaller clumps, aiding in subsequent stochastic mass growth in the cool phase \citep[e.g.,][]{Gronke2022}. The significance of this environmental effect on the formation of extended nebulae is supported by the fact that the nebulae in PKS0405$-$123, HE0238$-$1904, and TXS0206$-$048 are much larger in area than the nebulae in fields such as J0454$-$6116 and J2135$-$5316 where no massive close companions with consistent line-of-sight velocities were found in the nebulae footprint.  




\subsection{Energy transfer rate over seven decades in spatial scale}
\label{sec:epsilon}

\begin{figure}
	\includegraphics[width=\linewidth]{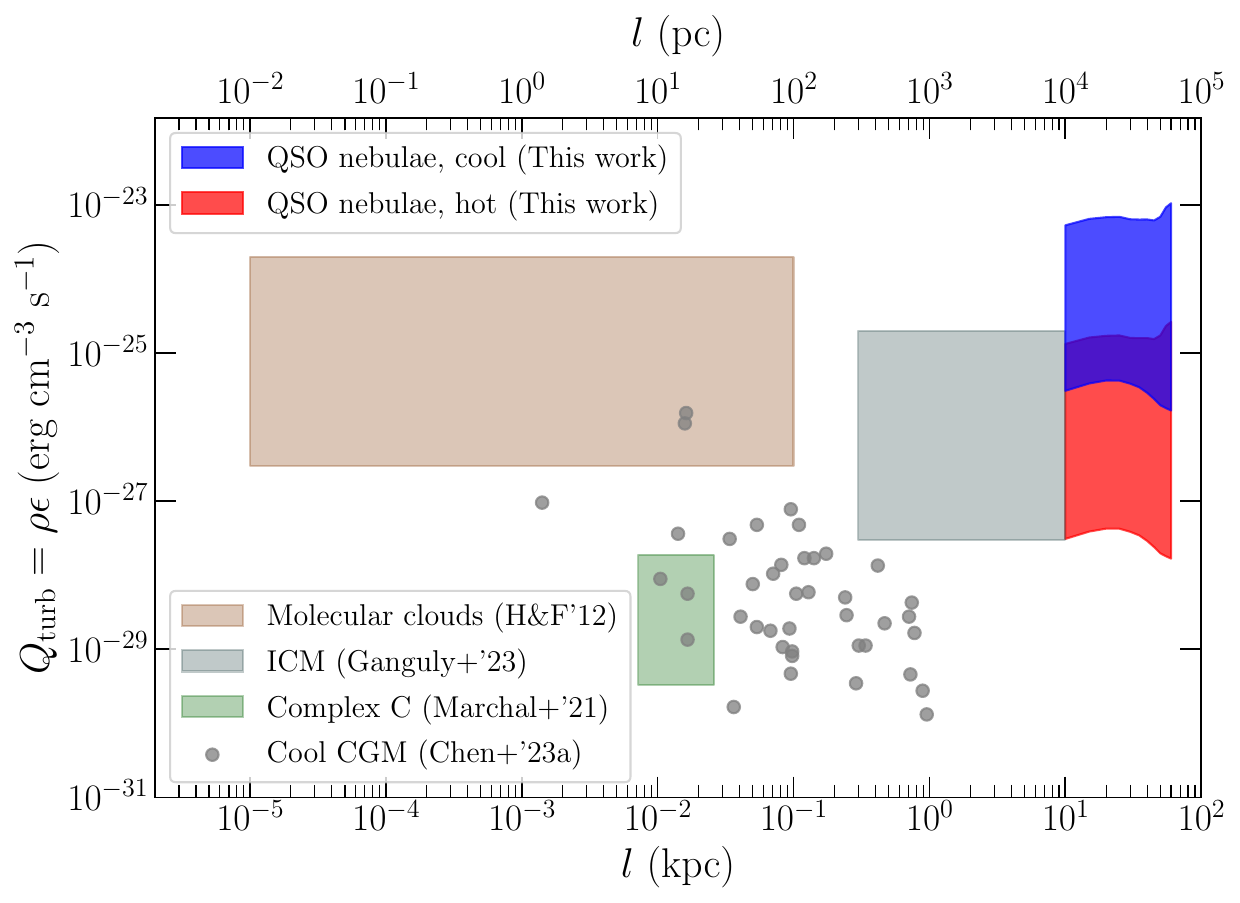}
    \caption{Turbulent heating rate $Q_{\rm turb}$ at different scales for different physical systems. The red and blue shaded regions show the estimated $Q_{\rm turb}$ for QSO nebulae at scales of $\approx 10$--60 kpc based on our sample.
    The calculations for the hot and cool gas phases assume densities of 0.01--1 cm$^{-3}$ and 1--40 cm$^{-3}$, respectively (see text).  The lower and upper bounds indicate the 16$^{\rm th}$--84$^{\rm th}$ percentile ranges for measurements across all eight nebulae. Measurements from \protect\cite{Ganguly2023} for ICM at scales $\approx 0.3$--10 kpc are shown by the gray shaded region.
    Results for star-forming molecular clouds at scales $\approx 0.01$--100 pc presented in \protect\cite{Hennebelle2012} are shown by the brown shaded region.
    \protect\cite{Marchal2021} measured $Q_{\rm turb}$ for a bright concentration location in the HVC Complex C at scales $\approx 6$--28 pc, as shown by the green shaded region.
    The gray points show the results for CGM cool clumps at scales $\approx 10$ pc -- 1 kpc based on absorption line measurements presented in \citet{Chen:2023b}. The turbulent heating rates in the QSO nebulae, the cool-core cluster ICM, and the star-forming molecular clouds are on average $\sim 1000$ times higher than that in Complex C and cool gas clumps probed in absorption. 
    } 
    \label{fig:rho_epsilon_compare}
\end{figure}

As described in \citetalias{Chen2023}, the energy transfer rate per unit mass $\epsilon$ can be calculated via the ``four-fifths law"\citep{Kolmogorov1941,Frisch1995}:
\begin{equation}
    \epsilon = \frac{5}{4}\left[\frac{|\langle \Delta v (r)^3\rangle|}{r}\right] \approx \frac{5}{4}\left[\frac{\langle |\Delta v (r)|^3\rangle}{r}\right].
\end{equation}
For Kolmogorov turbulence, $\epsilon$ is a constant at all scales within the inertial range.  For VSFs flatter (steeper) than the Kolmogorov expectation, the energy transfer rate would be higher (lower) on smaller spatial scales. 
Across different nebulae in our sample and on different scales between 10--60 kpc, the estimated $\epsilon$ shows a range of values between $\approx0.02$ cm$^2$ s$^{-3}$ and $\approx0.2$ cm$^2$ s$^{-3}$. For nebulae with both \oii and \oiii measurements, the values obtained using these two lines are consistent within uncertainty. This estimated range for $\epsilon$ with our sample is comparable to the measurements for H$\alpha$ filaments in core regions of nearby cool-core clusters \citep{Li2020,Ganguly2023} and molecular clouds in nearby \ion{H}{2} regions \citep[e.g.,][]{Hennebelle2012}.  Much lower estimates of $\epsilon\approx 10^{-7}$--10$^{-3}$ cm$^2$ s$^{-3}$ were obtained for CGM cool clumps probed through absorption line spectroscopy \citep[][]{Rauch2001,Chen:2023b}, and a Milky Way high-velocity cloud \citep[HVC;][]{Marchal2021}. 

To gain further insights into the differences between these dynamical systems, we convert the estimated $\epsilon$ to a turbulent heating rate per unit volume via $Q_{\rm turb}=\rho \epsilon$, where $\rho$ is the gas density and can span a wide range for gas in different phases. For the QSO nebulae in our sample, the \oii doublet line ratios suggest a median upper limit of gas density for the $T\!\sim\!10^4$ K cool phase of $\lesssim40$ cm$^{-3}$ \citep{Liu2023}, while an estimate of $\approx\!1$--5 cm$^{-3}$ is obtained assuming pressure equilibrium between typical AGN-illuminated [\ion{O}{2}]-emitting gas and the hot halo \citep{Johnson2022}.
Based on the [\ion{S}{2}]$\lambda\lambda6716,6731$ doublet ratio, observations of spatially extended nebula illuminated by the active galactic nucleus (AGN) in the Teacup galaxy at $z\!\sim\!0.1$ show that the gas density at distances of a few kpc away from the galaxy center is $\lesssim\!10$ cm$^{-3}$ \citep[][]{Venturi2023}.  Therefore, we adopt a range of 1--40 cm$^{-3}$ for the cool phase gas when calculating $Q_{\rm turb}$ to account for this wide range of uncertainty. For the hot phase with $T\!\approx\!10^6$--10$^7$ K, we adopt a density range of 0.01--1 cm$^{-3}$ \citep[e.g.,][]{Li2018}.  We obtain an estimated $Q_{\rm turb}$ of $\approx\!10^{-26}$--10$^{-22}$ erg cm$^{-3}$ s$^{-1}$ for the cool gas and $\approx\!10^{-28}$--10$^{-25}$ erg cm$^{-3}$ s$^{-1}$ for the hot gas, as shown by the blue and red shaded regions in Figure \ref{fig:rho_epsilon_compare}.  \cite{Ganguly2023} constrained the $Q_{\rm turb}$ of the ICM in the core regions of nearby cool-core clusters to be $\approx 10^{-28}$--10$^{-25}$ erg cm$^{-3}$ s$^{-1}$ (the gray shaded region in Figure \ref{fig:rho_epsilon_compare}), in agreement with our result for the hot phase.  For star-forming molecular clouds, measurements across a wide range of spatial scales of $\approx\!0.01$--100 pc led to an estimate of $Q_{\rm turb}\!\approx\!10^{-27}$--10$^{-24}$ erg cm$^{-3}$ s$^{-1}$ as presented in \cite{Hennebelle2012} and shown by the brown shaded region in Figure \ref{fig:rho_epsilon_compare}. \cite{Marchal2021} measured the density and kinematics of a bright concentration region near the edge of Complex C, an HVC in the Milky Way, which resulted in an estimated $Q_{\rm turb}\!\approx\!10^{-30}$--10$^{-28}$ erg cm$^{-3}$ s$^{-1}$ as shown by the green shaded region in Figure \ref{fig:rho_epsilon_compare}. Using non-thermal velocity widths of resolved absorption profiles and clump sizes inferred from photoionization models, 
\cite{Chen:2023b} 
constrained $Q_{\rm turb}$ to be $\approx\!10^{-30}$--10$^{-27}$ erg cm$^{-3}$ s$^{-1}$ for spectrally resolved cool clumps with a size scale of $\approx 10$ pc -- 1 kpc in the CGM. These are shown by the gray data points in Figure \ref{fig:rho_epsilon_compare}. 

It can be seen that the turbulent heating rates in the QSO nebulae, the cool-core cluster ICM, and the star-forming molecular clouds are on average $\sim 1000$ times higher than that in the MW HVC and cool gas clumps probed in absorption. Given that both Complex C and cool absorption clumps are expected to be in relatively quiescent, undisturbed environments \citep{Chen:2023b}, a possible explanation for this difference is that feedback due to star formation and AGN activities can significantly elevate the turbulent energy in the gaseous halos. However, caveats remain in this interpretation. As discussed in the previous section, the galaxy environments of the largest extended nebulae hint towards the scenario where tidal/merger interactions play a key role in stirring up the gas and facilitating the formation of multiphase structures, and the presence of a large amount of cool gas near the QSOs can lead to more efficient black hole accretion \citep[e.g.,][]{Prasad2015,Voit2017}. In this case, the elevated turbulent energy might be a precursor for fueling these luminous QSOs instead of a consequence of QSO feedback. 

For the first time, we are able to determine turbulent energy transfer rate in the diffuse cosmic gas over seven decades in spatial scale from $\sim 0.01$ pc to $\sim 100$ kpc, but the measurements rely on two distinct approaches at different spatial scales.  In particular, in the circumgalactic space, where we see three orders of magnitude difference in $Q_{\rm turb}$ from large to small scales, such distinction is also accompanied by differences in the way turbulence energy is determined. The gas turbulence probed in emission likely reflects the relative motions between different line-emitting clumps that trace the hot gas dynamics (as discussed in Section \ref{sec:discussion_dynamics}), while high-resolution absorption line studies likely probe turbulence internal to individual clouds. Therefore, the lack of overlapping spatial scales probed by emission and absorption prevents us from forming a consistent picture of turbulent energy cascade in galaxy halos, while systematic uncertainties remain when comparing turbulent flows based on VSF measurements and those from absorption-line analyses.   In \citetalias{Chen2023}, we discussed uncertainties associated with VSF measurements due to either projection effects \citep[see also e.g.,][]{vonHoerner1951,Xu2020} or PSF smoothing (see further discussion in \S\ \ref{sec:limitation_caveats}). While the smallest area accessible in emission measures is limited to the PSF size of the data, the absorption line technique averages cloud properties over the beam size that is dictated by the black hole accretion disk size (i.e., on the order of $\ll1$ pc). At the same time, absorption-line analyses are subject to uncertainties in the photo-ionizing background radiation field. 
Future observations using AO-assisted ground-based IFSs and/or space-based IFSs can extend the small scales probed in the VSFs to $\lesssim 10$ kpc for the line-emitting gas, bridging the gap in spatial scales accessible between emission and absorption studies. A sample of systems with both extended line emission and high-resolution absorption line data will also greatly aid in the investigation of this discrepancy in $Q_{\rm turb}$. 




Finally, we note that these measurements of turbulent motions in QSO nebulae imply that turbulence is insufficient in providing the required energy to offset cooling at tens of kpc scales in the QSO environments. In Section 5.1 of \citetalias{Chen2023}, we compared the turbulent heating rate and the radiative cooling rate, utilizing measurements from TXS0206$-$048. The calculations considered the gas mass within a 50 kpc radius of a $5\times10^{13}\,\msun$ halo, assuming an NFW mass profile and that gas of all phases is perfectly coupled dynamically. This approximate evaluation shows that the turbulent heating rate is on par with the luminosity of [\ion{O}{2}] or [\ion{O}{3}], yet it constitutes only approximately 0.05\% of the QSO bolometric luminosity.

\subsection{Velocity dispersion along the line of sight versus in the plane of the sky}
\label{sec:velocity_dispersion}
\begin{figure}
	\includegraphics[width=\linewidth]{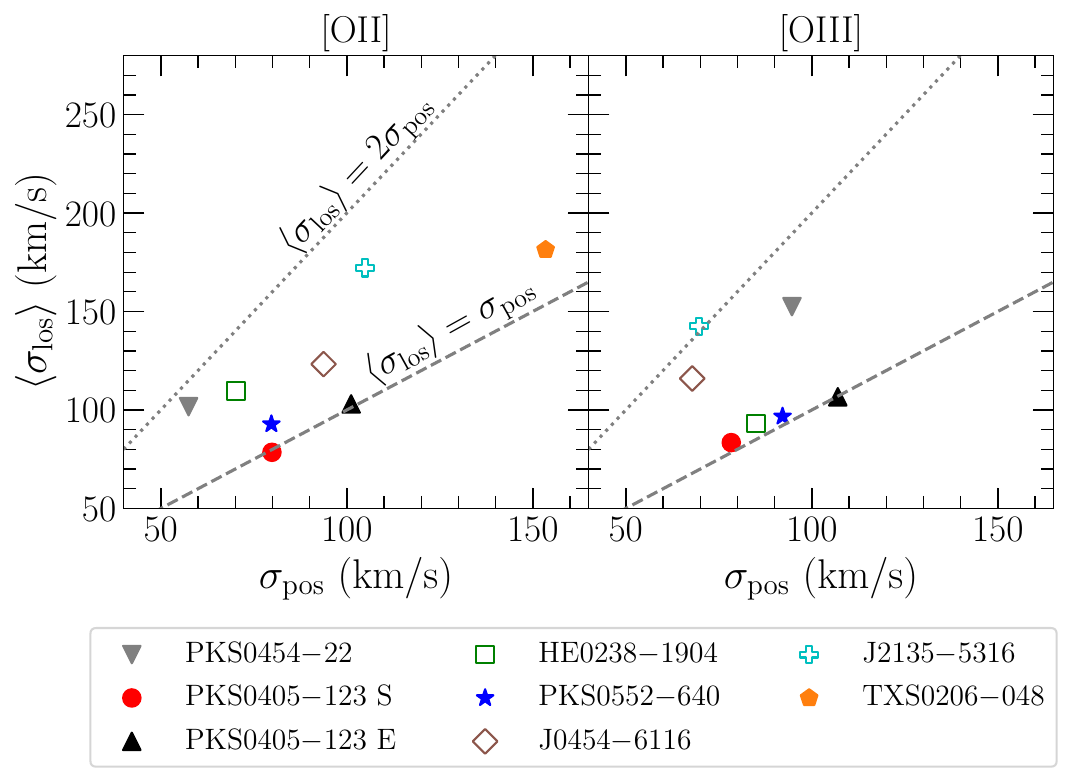}
    \caption{The positional velocity dispersion in the plane of the sky, $\sigma_{\rm pos}$, versus the mean velocity dispersion along the line of sight, $\langle\sigma_{\rm los}\rangle$.  The left and right panels show measurements using the \oii and \oiii emission lines, respectively.  The dashed line shows the relation $\sigma_{\rm pos}=\langle\sigma_{\rm los}\rangle$, and the dotted line indicates the relation where $\langle\sigma_{\rm los}\rangle$ is twice the value of $\sigma_{\rm pos}$. For all nebulae in our sample, $\sigma_{\rm pos}\lesssim\langle\sigma_{\rm los}\rangle$.}
    \label{fig:sigma_los_vs_pos}
\end{figure}

In Figure \ref{fig:sigma_los_vs_pos}, we show the velocity dispersion in the plane of the sky, $\sigma_{\rm pos}$, versus the mean velocity dispersion along the line of sight, $\langle\sigma_{\rm los}\rangle$.  $\sigma_{\rm pos}$ is quantified as the standard deviation of the line-of-sight velocity from spaxels included in the VSF measurements (see discussion in Section \ref{sec:vsf_measurements} and the velocity maps in Figures \ref{fig:PKS0405_seg1_o2}--\ref{fig:PKS0552_o3}), and $\langle\sigma_{\rm los}\rangle$ is the mean line width (obtained through a single-component Gaussian fit) for the same set of spaxels.  We show results for both \oii and \oiii emission as they can differ in $\sigma_{\rm pos}$ and $\langle\sigma_{\rm los}\rangle$ due to the different footprints of the two lines. The statistical uncertainties of both velocity dispersions estimated through Monte Carlo resampling are small and are not shown in Figure \ref{fig:sigma_los_vs_pos}.  The measurements for $\sigma_{\rm pos}$ before and after removing the uni-directional velocity gradient in the plane of the sky are consistent with each other to within $\approx 20$ km/s.  Therefore, for clarity, we only show the values obtained using the directly measured [\ion{O}{2}] and [\ion{O}{3}] velocity maps.  

It can be seen that for all nebulae in our sample, $\sigma_{\rm pos}\lesssim\langle\sigma_{\rm los}\rangle$. This observation agrees with the general trend seen in spatially-resolved data for \ion{H}{2} regions where the velocity dispersion along the line of sight exceeds the velocity dispersion in the plane of the sky \citep[e.g.,][]{Lagrois2011,Arthur2016,Garcia-Vazquez2023}. One possible explanation for this trend is the smoothing effect due to multiple line-emitting clouds along the line of sight contributing to the observed velocity centroid, leading to reduced velocity dispersion in the plane of the sky. In addition, the contribution from bulk/coherent motions along the line of sight will also result in larger $\sigma_{\rm los}$. 
To investigate this possibility, we adopt a simple assumption that $\langle \sigma_{\rm los}\rangle^2 = [\sigma_{\rm pos}^2 + (v_{\rm grad, los}\times L_{\rm los})^2]$, where $v_{\rm grad, los}$ is the velocity gradient along the line of sight.  We approximate the depth of the nebula $L_{\rm los}$ to be the square root of the nebula size (see Table \ref{tab:emission_line_properties}), and derive a velocity gradient of $v_{\rm grad, los}\approx 0.5$--3 km/s/kpc for different nebulae.  The range of this derived $v_{\rm grad, los}$ is in qualitative agreement with the best-fitting velocity gradient in the plane of the sky (see Table \ref{tab:slopes}), suggesting that bulk flows 
along the line of sight may be non-negligible. 
In contrast, the velocity dispersion across the plane of the sky provides a robust tracer of the underlying velocity variance at scales $\gtrsim10$ kpc, particularly when a credible model for the coherent shear in the plane of the sky can be obtained with the spatially-resolved velocity measurements, as pointed out by previous studies  \citep[e.g.,][]{Stewart2022, Garcia-Vazquez2023}.


\subsection{Power-law turnover scale for the VSFs}
\label{sec:turnover_correlation}

\begin{figure}
	\includegraphics[width=\linewidth]{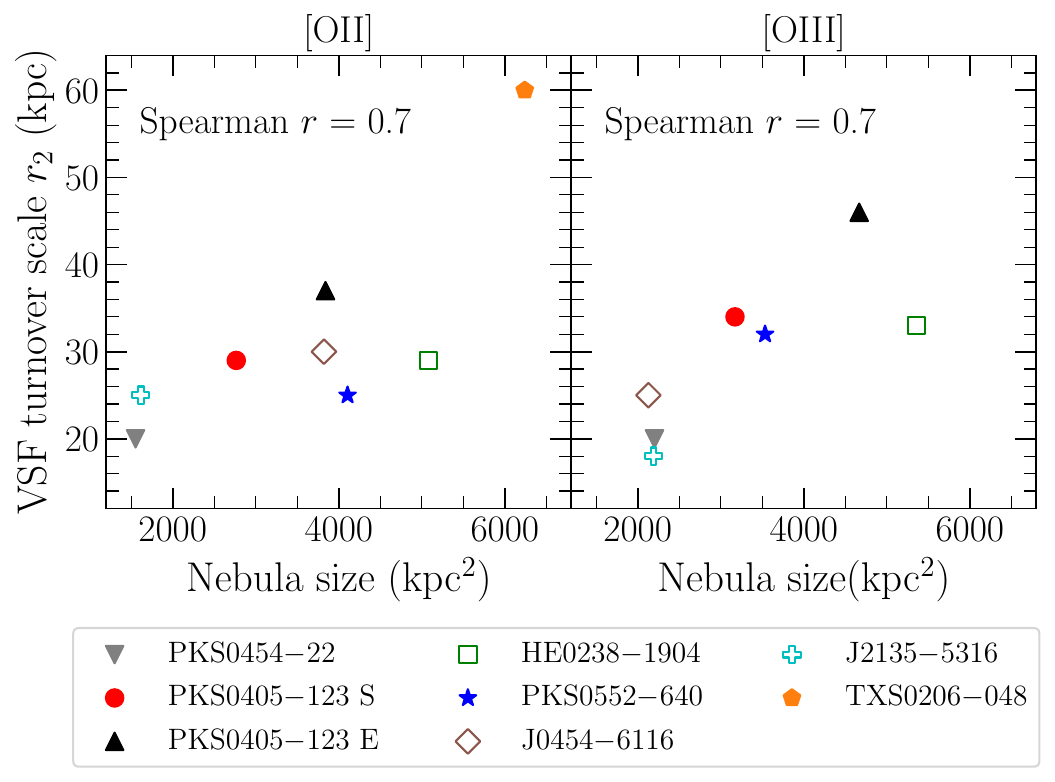}
    \caption{Nebula size versus the VSF turnover scale $r_2$ for all eight nebulae in the sample.  The left and right panels show the values using the \oii and \oiii emission lines, respectively. There is a moderate correlation between nebula size and the VSF turnover scale. }
    \label{fig:size_turnover_correlation}
\end{figure}

As discussed in Section \ref{sec:s2_slopes}, the shapes of the VSFs generally do not follow a single power-law across the entire range of scales probed.  While additional structures in the VSFs may provide hints for different physical processes present in the nebulae, we caution that the limited nebula size and signal-to-noise can hinder a robust interpretation of these structures.

In particular, we note that there is a moderate correlation (with a Spearman's $r$ coefficient of 0.7) between the VSF turnover scale $r_2$ (see Section \ref{sec:s2_slopes}) and the size of the nebula for both the \oii and \oiii emission, as shown in Figure \ref{fig:size_turnover_correlation}.  This correlation indicates that the deviation of the VSF from a single power-law at larger scales is in part due to the limited nebula size probed by the data given the detection limit. Previous studies have also shown that boundaries of clouds/nebulae can artificially flatten the VSFs at large scales that mimic the signature of energy injection and affect the interpretation of the data \citep[e.g.,][]{Ganguly2023,Garcia-Vazquez2023}.  In addition, the smooth transition between the inertial range and the energy injection scale can cause the VSF slopes to taper off at a scale as small as half of the true energy injection scale \citep[][]{Federrath2021} and further complicate the interpretation of a flattening signal in the VSFs. 

\begin{figure}
	\includegraphics[width=\linewidth]{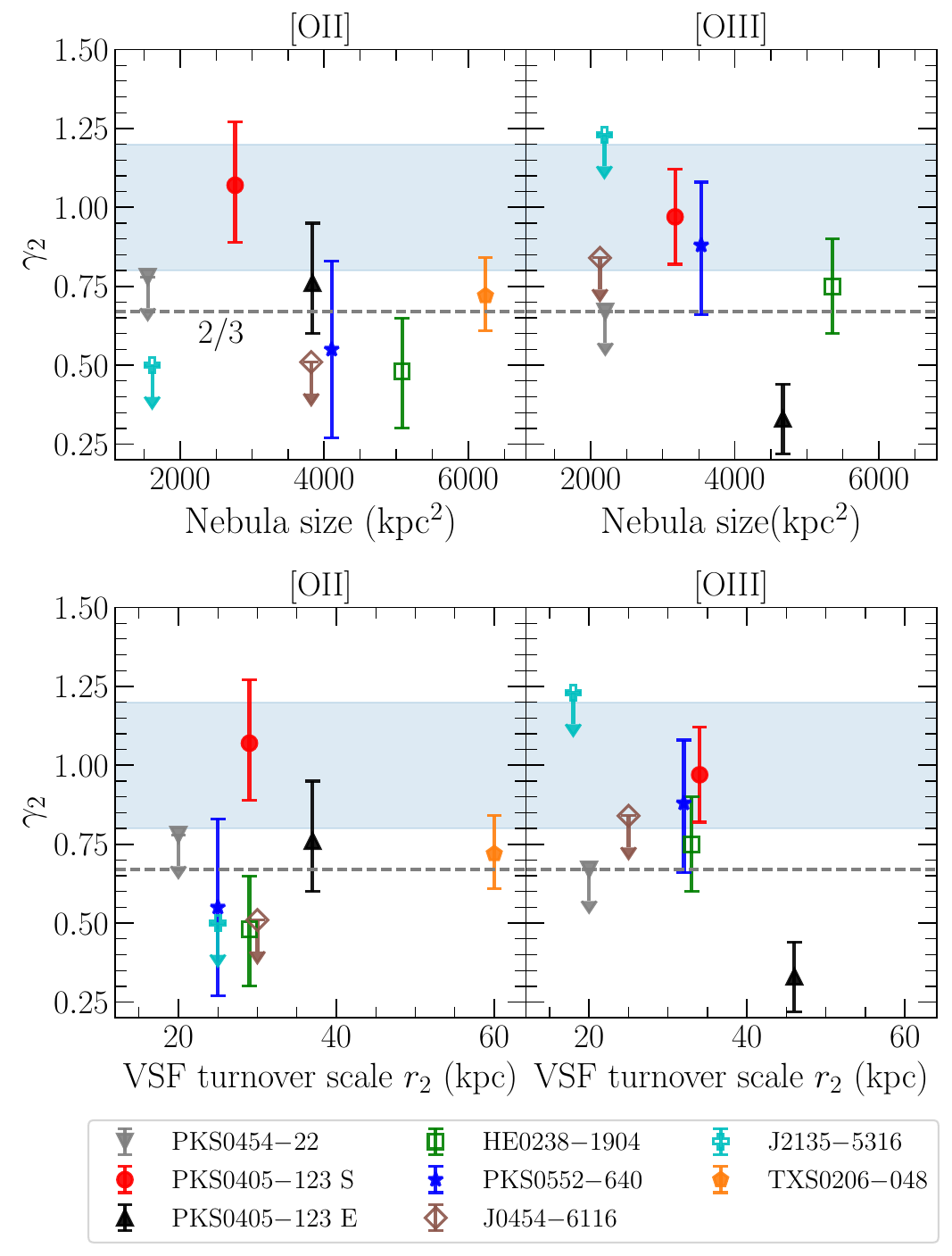}
    \caption{{\it Top row: }Nebula size versus the 2nd-order VSF slope $\gamma_2$.  {\it Bottom row: }The VSF turnover scale $r_2$ versus the 2nd-order VSF slope $\gamma_2$.  The results based on the \oii measurements are shown in the left panels while the results from \oiii are shown in the right panels.  The horizontal blue shaded regions mark the measurements for local \hii regions presented in \protect\cite{Garcia-Vazquez2023}, which are on average higher than the slopes constrained for QSO nebulae. No discernible correlations are found between the 2nd-order power-law slopes ($\gamma_2$) and VSF turnover scale ($r_2$) or nebulae size.}
    \label{fig:size_turnover_vs_gamma2}
\end{figure}

Given the abovementioned caveats, we refrain from interpreting $r_2$ or VSF flattening scales in our sample as indicative of energy injection scales. However, Figure \ref{fig:size_turnover_vs_gamma2} indicates no discernible correlations between the constrained 2nd-order power-law slopes ($\gamma_2$) and VSF turnover scale ($r_2$) or nebula size, underscoring the robustness of $\gamma_2$ measurements. Measurements from local \ion{H}{2} regions reported by \cite{Garcia-Vazquez2023} result in larger $\gamma_2$ values on average (shown in the blue shaded region in Figure \ref{fig:size_turnover_vs_gamma2}), suggesting elevated Mach numbers in local \ion{H}{2} regions and/or increased susceptibility to projection smoothing in their observations (for more discussion of projection effects see Section \ref{sec:limitation_caveats} below).

\subsection{Limitations and caveats}
\label{sec:limitation_caveats}
A notable limitation in the present study arises from the projection effect inherent in the data. Several studies have investigated how VSFs are affected by the use of projected measurements. Analytically, \cite{vonHoerner1951} derived that for volume-filling gas, the projection effect depends on the spatial scales probed: VSFs are steepened when measuring separation scales smaller than the depth of the cloud along the line of sight, while the VSF slopes recover to the intrinsic value at scales exceeding the cloud depth. This result is sometimes referred to as the ``projection smoothing" effect and was independently confirmed by \cite{Odell1987} and \cite{Xu2020}. On the other hand, \cite{Zhang2022} used numerical simulations to show that for spatially confined structures (e.g., isolated filaments), the projection effect flattens the VSFs. As we have discussed in Section \ref{sec:discussion_dynamics}, the dynamical state of the nebulae examined in this work indicates that the cool line-emitting gas is embedded in the hot ambient medium and traces the turbulent motions of the hot, volume-filling gas. Therefore, our measurements are more likely affected by the ``projection smoothing" effect, suggesting that the intrinsic VSF slopes may be flatter than the values reported in Table \ref{tab:slopes}, which still supports our interpretation of the subsonic/transonic gas motions. In addition to whether the gas is volume-filling or spatially confined, in reality, the projection effect will also depend on detailed properties of the system such as density/emissivity fluctuations and the three-dimensional geometry of the gas structure. Detailed investigations using high-resolution numerical simulations are needed to robustly quantify and calibrate the projection effect in more realistic environments. 

Another main limitation of the current study is the restricted dynamic range in the VSF measurements, which is confined to approximately one decade or less in projected distance separation. This restriction prevented us from obtaining robust constraints on the VSFs slopes for several systems in our sample.  While the largest separation is determined by the nebula size given the detection threshold, the smallest separation accessible in the data is dictated by the spatial sampling (i.e., angular size per spatial pixel) as well as the PSF size. As ground-based observations without adaptive optics (AO) are fundamentally limited by atmospheric seeing, improving the dynamic range towards small scales requires conducting AO-assisted observations on the ground (e.g., with VLT/ERIS in the infrared and using the Narrow-Field-Mode on VLT/MUSE in the optical) with longer exposure times to reach sufficient signal-to-noise. Alternatively, space-based IFSs such as {\it JWST}/NIRSpec with unprecedented spatial resolution have also started delivering an increasing sample of spatially-resolved observations of the CGM \citep[e.g.,][]{Wylezalek2022,Veilleux2023}. Finally, with a fixed PSF size, targeting systems at lower redshifts with a higher angular-to-physical size ratio can also help increase the VSF dynamic range. However, few extended ($\gtrsim50$ kpc) nebulae have been discovered at $z<0.5$ \citep[e.g.,][]{Rupke2019,Chen2019,Venturi2023} and additional effort is required to expand the sample size of low-redshift extended nebulae. 


\section{Conclusion}
\label{sec:conclusion}
This paper presents an ensemble study of the turbulent motions in eight extended nebulae surrounding seven QSOs at $z\approx0.5$--1.1.  Using the \oii and/or \oiii emission lines, we measure the line-of-sight velocity fields and construct the velocity structure functions (VSFs). We probed the dynamical state of the gas illuminated by the QSO radiation field at scales $\approx10$--100 kpc. Our main conclusions are:
\begin{itemize}
    \item Five out of the eight nebulae in our sample have a constrained power-law slope of the 2nd-order VSFs, $\gamma_2$, between $\approx0.3$--1.1, while the other three nebulae have loose constraints corresponding to 95\% upper limits of $\lesssim 0.5$--1.5, as shown in Figures \ref{fig:S2} and \ref{fig:size_turnover_vs_gamma2} and discussed in Section \ref{sec:s2_slopes}.  To within the 2-$\sigma$ measurement uncertainty, the slopes are either consistent with the expectation from Kolmogorov turbulence or flatter, suggesting that the gas motions are subsonic. 
    
    \item Removing a best-fitting unidirectional velocity gradient from the line-of-sight velocity maps flattens the VSFs in general, but also leads to larger uncertainties due to a reduced dynamic range in the VSFs that can be used for a single power-law fit.  The results before and after removing a velocity gradient are consistent within the range of the uncertainty, as shown in Figure \ref{fig:S2}. 
    
    \item Complementing the measurements for the 2nd-order VSF slopes, $\gamma_2$, the ESS slope ratios $\gamma_p/\gamma_3$ for $p=1$--6 are also in agreement with the expectation of subsonic turbulence, as shown in Figure \ref{fig:ess} and discussed in Section \ref{sec:ess_slopes}.  The only exception is the nebula surrounding the QSO field HE0238$-$1904, where the $\gamma_p/\gamma_3$ ratios are consistent with the supersonic MHD turbulence prediction by \cite{Boldyrev2002} both before and after removing a uni-directional gradient field.  A more detailed investigation of this field and a larger sample size are required to shed light on whether this field is a special case. 
    
    \item The subsonic motions in the QSO nebulae suggest that the line-emitting cool clouds with $T\sim 10^4$ K are embedded within a hot ambient medium with $T\sim 10^6$--10$^7$ K.  Adopting the sound speed of the hot medium of $c_{\rm s, hot}\approx$500 km/s, we estimate the Mach number of the cool clouds to be $\approx 0.2$--0.5, consistent with the observed VSF properties. The subsonic nature of gas motions supports a scenario where the cool clumps condense out of the hot gas, carrying the turbulent memory of the hot halo and serving as tracers of hot phase dynamics (see Section \ref{sec:discussion_dynamics}). 
    
    \item No discernible differences are seen in VSF properties between radio-loud and radio-quiet QSO fields, suggesting that the collimated jets and their inflated bubbles do not play a critical role in shaping the dynamical state of the gas on $\sim$tens of kpc scales.
    
    \item Comparing the mean velocity dispersion along the line of sight, $\langle\sigma_{\rm los}\rangle$, and the velocity dispersion observed in the plane of the sky,  $\sigma_{\rm pos}$, we find that $\langle\sigma_{\rm los}\rangle\gtrsim\sigma_{\rm pos}$ for all fields (Figure \ref{fig:sigma_los_vs_pos}).  We discuss that projection effects and bulk motion along the line of sight are possible sources for the larger dispersion (see Section \ref{sec:velocity_dispersion}).

    \item The turbulent heating rate per unit volume, $Q_{\rm turb}$, in the QSO nebulae is estimated to be $\sim 10^{-26}$--$10^{-22}$ erg cm$^{-3}$ s$^{-1}$ for the cool phase and $\sim 10^{-28}$--$10^{-25}$ erg cm$^{-3}$ s$^{-1}$ for the hot phase at scales $\approx 10$--60 kpc.  This range is in agreement with the measurements for intracluster medium and star-forming molecular clouds but is $\sim 1000$ times higher than that estimated for Milky Way Complex C and cool circumgalactic gas clumps probed in low-ion absorption lines, as shown in Figure \ref{fig:rho_epsilon_compare} and discussed in Section \ref{sec:epsilon}. While the difference in $Q_{\rm turb}$ might be a signpost for AGN/stellar feedback, a robust investigation into the systematics of the different measurements is required to shed light on this discrepancy.

\end{itemize}
Future observations of extended nebulae using AO-assisted IFSs on the ground (e.g., MUSE Narrow-Field-Mode) and/or space-based IFSs (e.g., {\it JWST}/NIRSpec IFU) will help extend the small scales probed in VSFs to $\lesssim10$ kpc, improving the robustness of the VSF constraints and bridging the gap between $Q_{\rm turb}$ measured by emission and absorption techniques. The findings of this ensemble study align with the recent emerging picture of the multiphase CGM where different gas phases are intricately connected throughout their formation and evolution history. Turbulence plays a critical role in facilitating nonlinear interactions within the gaseous halos, which in turn promote further developments of turbulence. For shaping the dynamical properties of gas traced by [\ion{O}{2}] and [\ion{O}{3}] at scales $\gtrsim10$ kpc, environmental effects (e.g., tidal interactions, galaxy mergers, gas accretion) may dominate over QSO feedback. These findings can be directly compared with high-resolution numerical simulations to shed light on detailed physical mechanisms that govern the driving and development of turbulence in the CGM. 


\begin{acknowledgments}
We thank Fausto Cattaneo and Jenny Greene for helpful discussions throughout this work. We also thank Yuan Li for constructive feedback on the discussions of this paper. 
HWC and MCC acknowledge partial support from NSF AST-1715692 grants. ZQ acknowledges partial support from NASA ADAP grant 80NSSC22K0481. 
JIL is supported by the Eric and Wendy Schmidt AI in Science Postdoctoral Fellowship, a Schmidt Futures program.
SC gratefully acknowledges support from the European Research Council (ERC) under the European Union’s Horizon 2020 Research and Innovation programme grant agreement No 864361. FSZ acknowledges the support of a Carnegie Fellowship from the Observatories of the Carnegie Institution for Science.
EB acknowledges support by NASA under award number 80GSFC21M0002. 
This research has made use of the services of the ESO Science Archive Facility and the Astrophysics Data Service (ADS)\footnote{\url{https://ui.adsabs.harvard.edu/classic-form}}. The analysis in this work was greatly facilitated by the following \texttt{python} packages:  \texttt{Numpy} \citep{Numpy}, \texttt{Scipy} \citep{Scipy}, \texttt{Astropy} \citep{astropy:2013,astropy:2018}, \texttt{Matplotlib} \citep{Matplotlib}, and \texttt{MPDAF} \citep{MPDAF}.  This work was completed with resources provided by the University of Chicago Research Computing Center.
\end{acknowledgments}

%






\appendix

\section{Absence of the luminosity--velocity dispersion relation}
\label{sec:appendix_l_sigma}
\begin{figure}
	\includegraphics[width=0.45\linewidth]{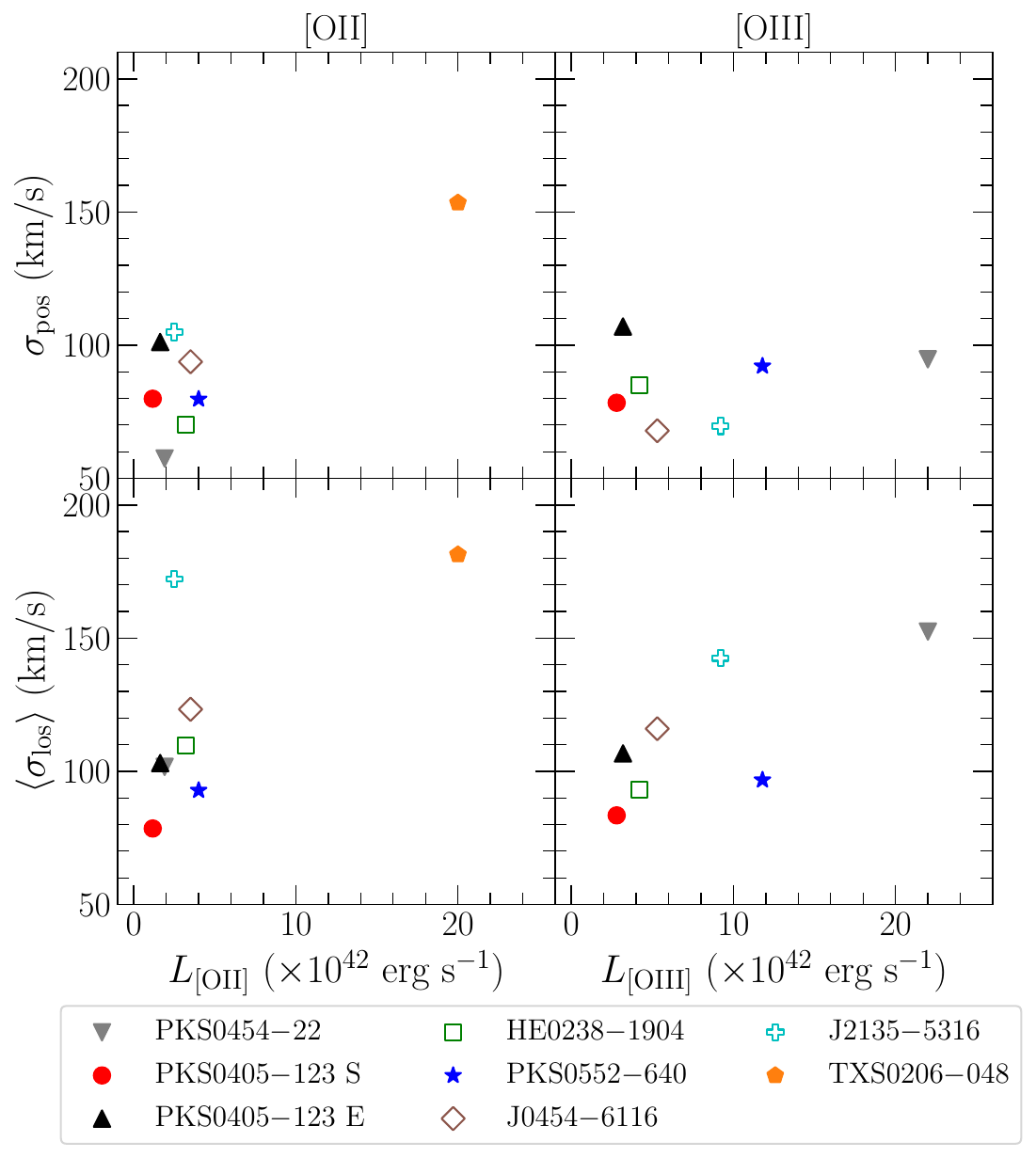}
    \caption{The nebula emission line luminosity (for both \oii and \oiii) versus the velocity dispersion, both along the line of sight and in the plane of the sky. We do not observe a $L-\sigma$ correlation using the [\ion{O}{2}] and [\ion{O}{3}] extended emission surrounding QSOs. }
    \label{fig:sigma_vs_lum}
\end{figure}

For local \hii regions as well as \hii galaxies (at both low and high redshifts), a $L$-$\sigma$ relation corresponding to the correlation between the luminosity of the region/galaxy in a certain line emission (such as H$\alpha$ and H$\beta$) and its velocity dispersion is commonly observed \citep[e.g.,][]{Melnick1987, Gonzalez-Moran2021}. In our QSO nebulae sample, however, we do not observe such a correlation, as shown in Figure \ref{fig:sigma_vs_lum}. The contrast here likely arises from the different emission mechanisms for recombination lines versus collisionally-excited lines, the former of which is more well coupled to the total mass and stellar feedback in the \hii regions/galaxies. In addition, QSOs are variable and the number of ionization photons output by QSOs is subject to significant changes on timescales of $\lesssim$tens of Myr \citep[e.g.,][]{Schawinski2015,Sun2017,Shen2021}{}{}, further weakening a correlation between the luminosity of the surrounding nebulae and the velocity dispersion of the gas. 

\section{Measurements for individual nebulae}
\label{sec:appendix_individual_nebulae}
Here we present the VSFs measurements for individual nebulae in PKS0405$-$123, PKS0552$-$640, and HE0238$-$1904.  The measurements for PKS0454$-$22, J0454$-$6116, and J2135$-$5316 can be found in the Appendix of \citetalias{Chen2023}. 

\begin{figure*}
	\includegraphics[width=\textwidth]{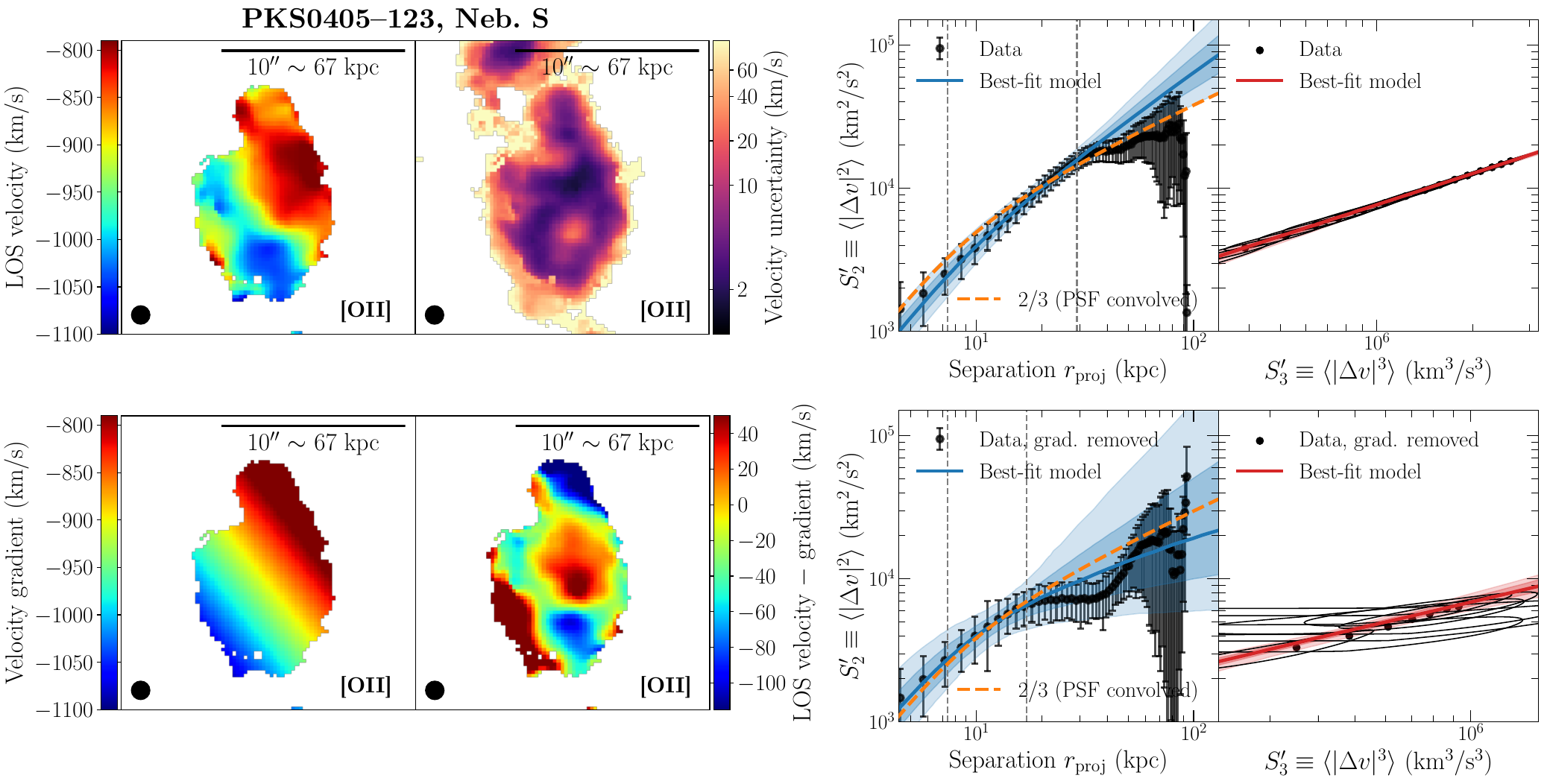}
    \caption{{\it Left-hand panels:} The best-fitting line-of-sight velocity map (top left), the line-of-sight velocity uncertainty (top right), the best-fitting uni-directional velocity gradient map (bottom left), and the residual velocity map after removing the velocity gradient from the line-of-sight velocity map (bottom right) for the southern nebula around PKS0405$-$123 based on the \oii emission.  {\it Right-hand panels:} The observed 2nd-order VSF, $S_2^\prime$, as well as the ESS correlation between $S_2^\prime$ and $S_3^\prime$.  The top row shows the measurements using the best-fitting line-of-sight velocity map shown in the top left panel, and the bottom row shows the results after removing a uni-directional gradient, as shown in the bottom right panel on the left. }
    \label{fig:PKS0405_seg1_o2}
\end{figure*}

\begin{figure*}
	\includegraphics[width=\textwidth]{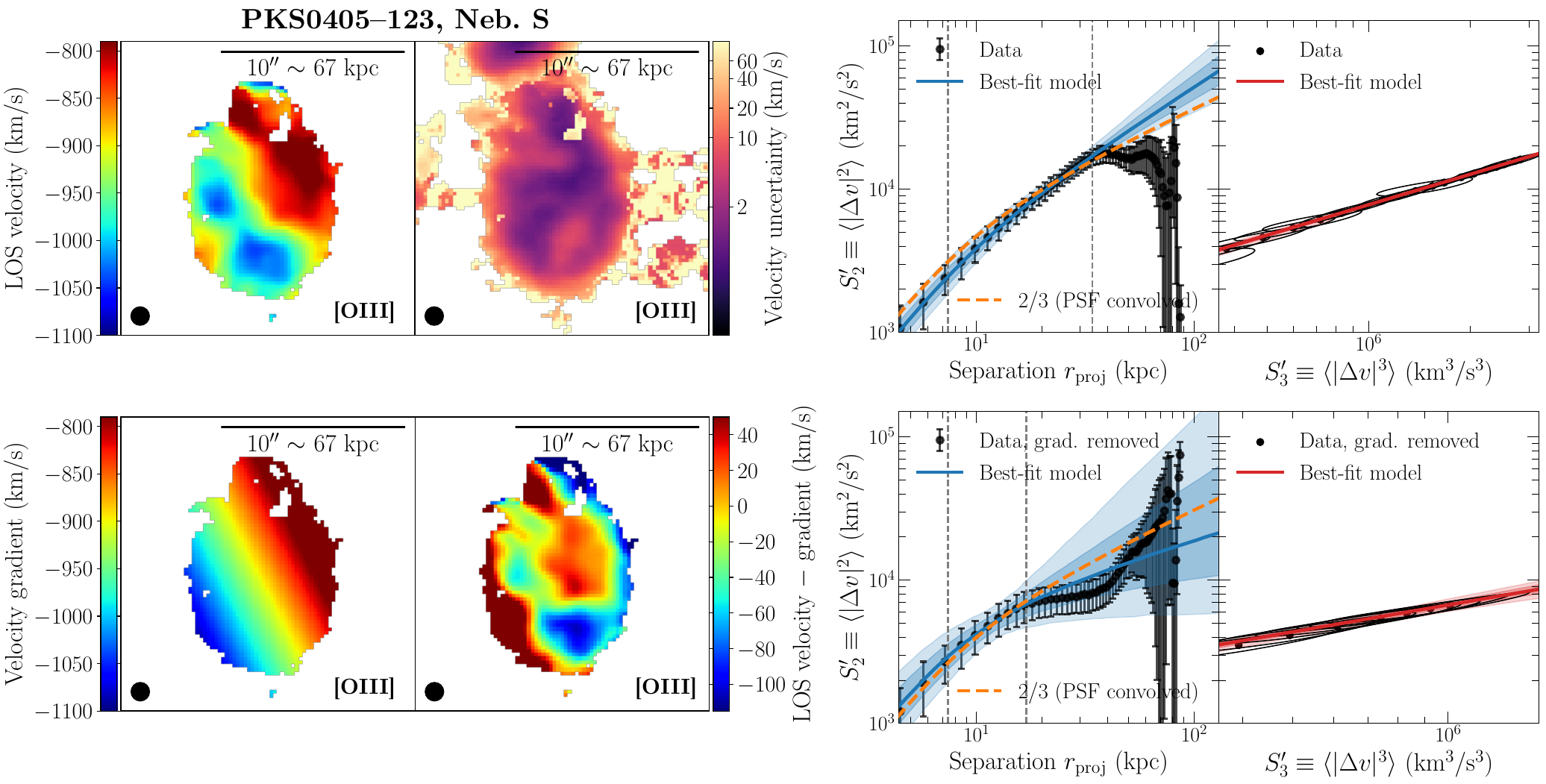}
    \caption{Same as Figure \ref{fig:PKS0405_seg1_o2}, but for the southern nebula around PKS0405$-$123 based on the \oiii emission.}
    \label{fig:PKS0405_seg1_o3}
\end{figure*}

\begin{figure*}
	\includegraphics[width=\textwidth]{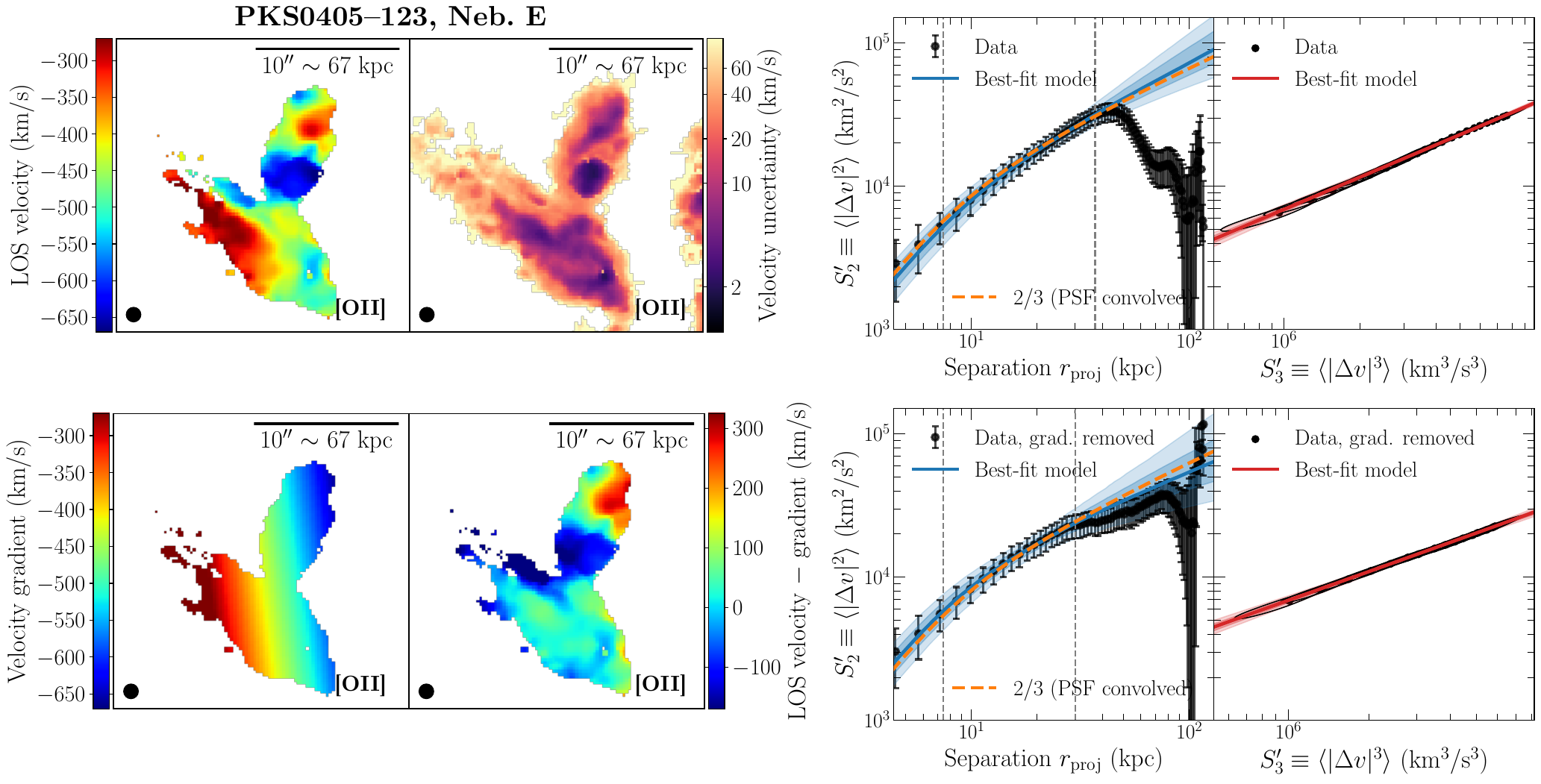}
    \caption{Same as Figure \ref{fig:PKS0405_seg1_o2}, but for the eastern nebula around PKS0405$-$123 based on the \oii emission.}
    \label{fig:PKS0405_seg2_o2}
\end{figure*}

\begin{figure*}
	\includegraphics[width=\textwidth]{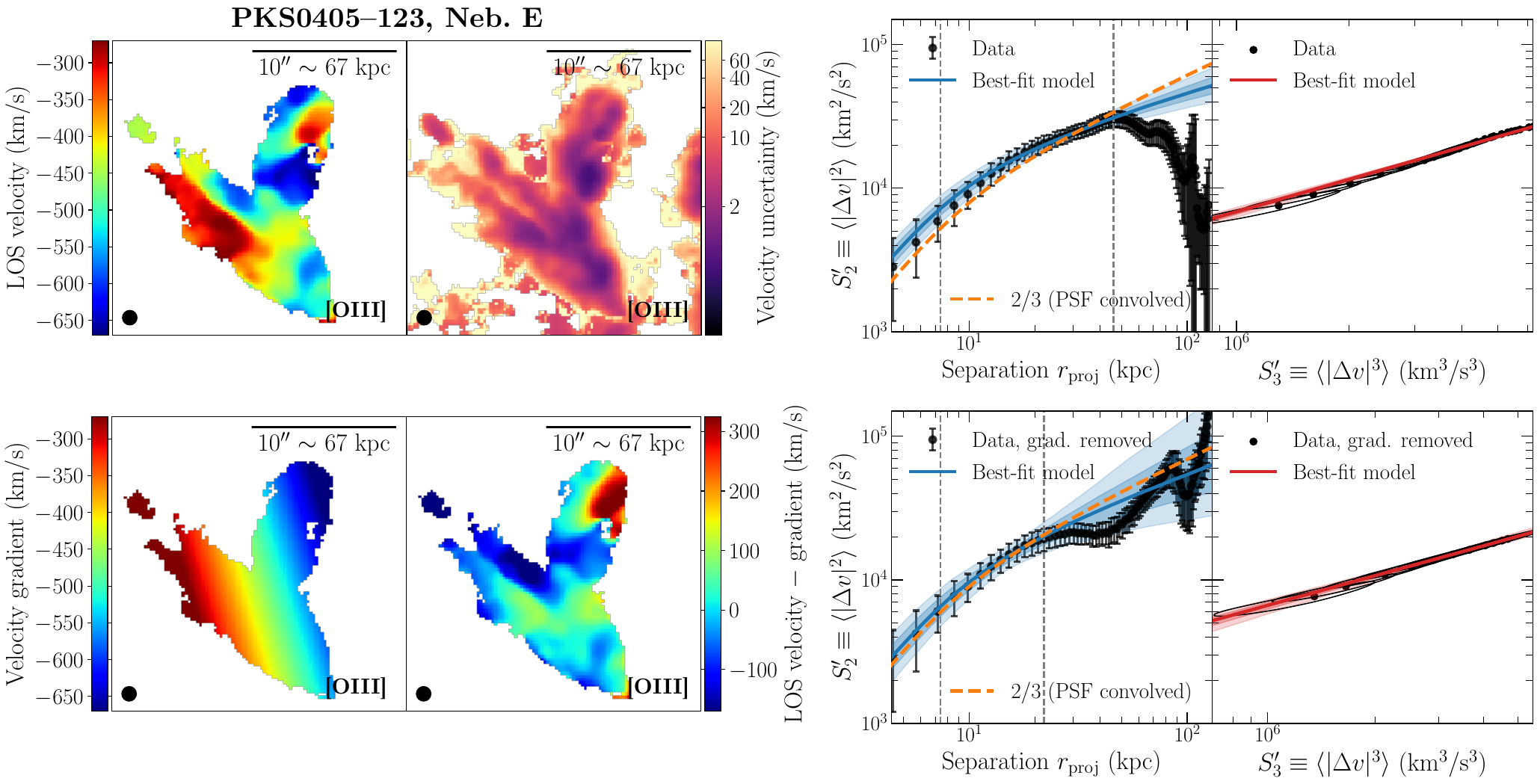}
    \caption{Same as Figure \ref{fig:PKS0405_seg1_o2}, but for the eastern nebula around PKS0405$-$123 based on the \oiii emission.}
    \label{fig:PKS0405_seg2_o3}
\end{figure*}

\begin{figure*}
	\includegraphics[width=\textwidth]{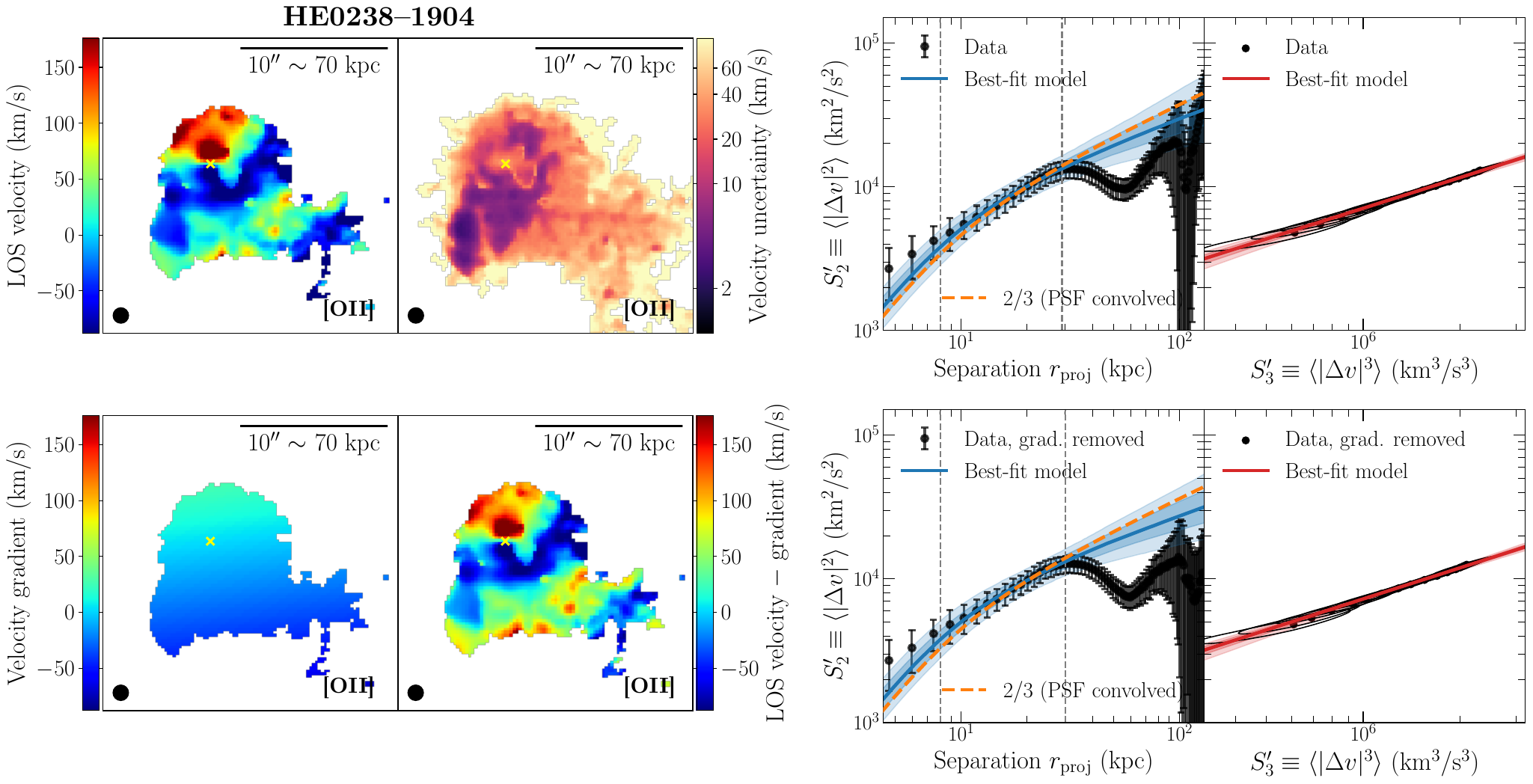}
    \caption{Same as Figure \ref{fig:PKS0405_seg1_o2}, but for the field of HE0238$-$1904 based on the \oii emission.}
    \label{fig:HE0238_o2}
\end{figure*}

\begin{figure*}
	\includegraphics[width=\textwidth]{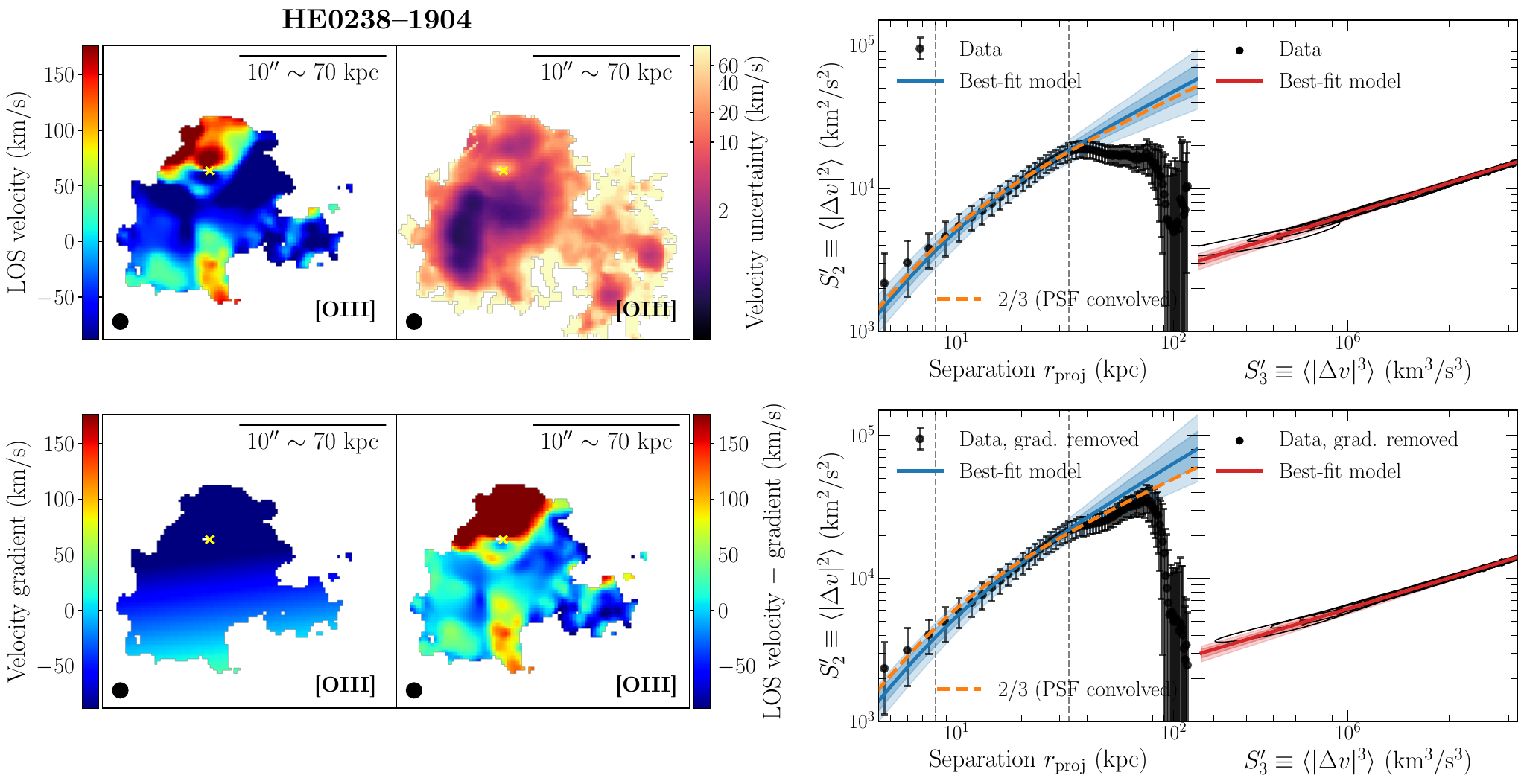}
    \caption{Same as Figure \ref{fig:PKS0405_seg1_o2}, but for the field of HE0238$-$1904 based on the \oiii emission.}
    \label{fig:HE0238_o3}
\end{figure*}

\begin{figure*}
	\includegraphics[width=\textwidth]{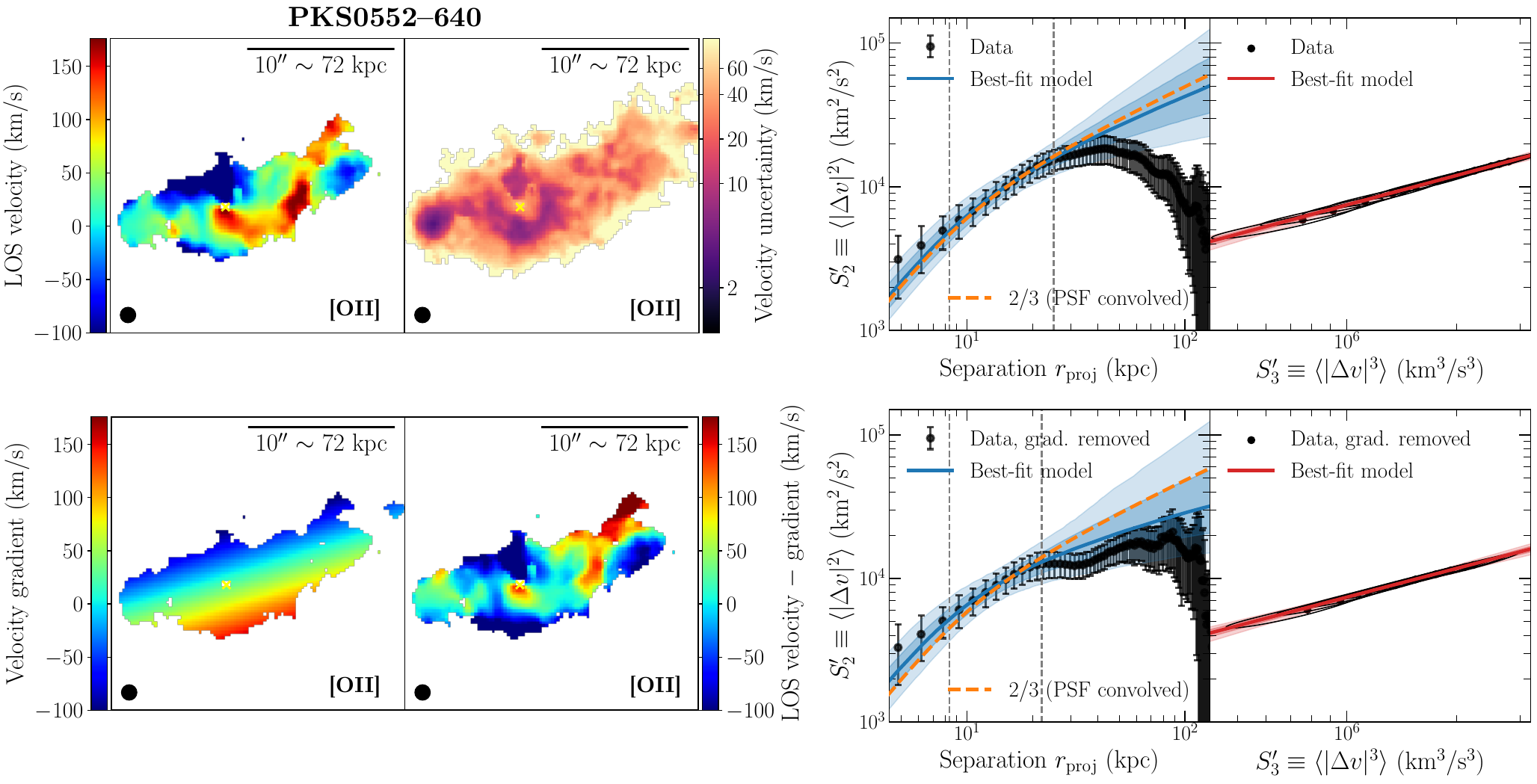}
    \caption{Same as Figure \ref{fig:PKS0405_seg1_o2}, but for the field of PKS0552$-$640 based on the \oii emission.}
    \label{fig:PKS0552_o2}
\end{figure*}

\begin{figure*}
	\includegraphics[width=\textwidth]{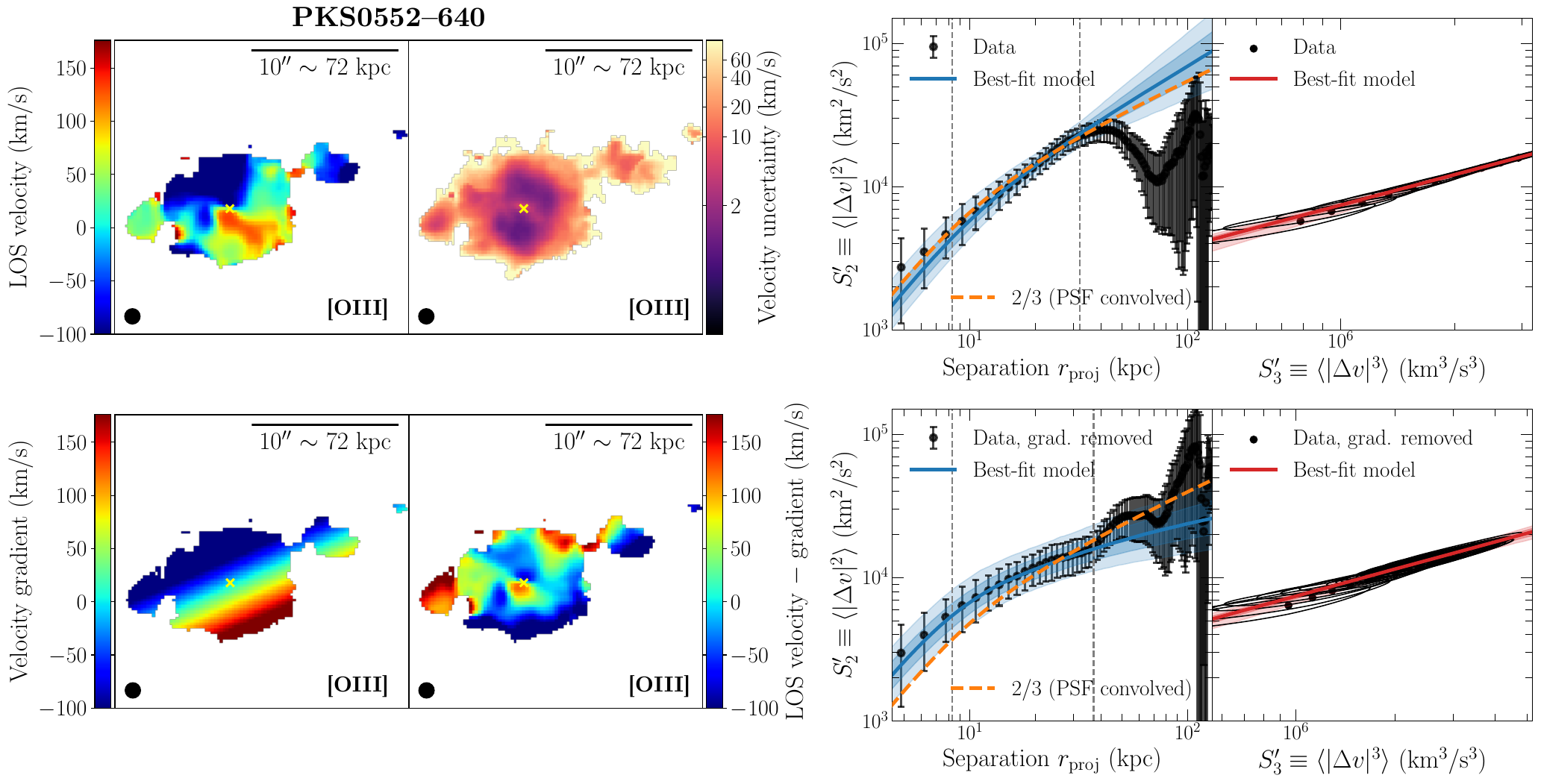}
    \caption{Same as Figure \ref{fig:PKS0405_seg1_o2}, but for the field of PKS0552$-$640 based on the \oiii emission.}
    \label{fig:PKS0552_o3}
\end{figure*}


\clearpage

\bibliography{main}{}
\bibliographystyle{aasjournal}



\end{document}